\documentstyle[12pt,psfig]{article}

\makeatletter

\long\def\@caption#1[#2]#3{\addcontentsline{\csname
  ext@#1\endcsname}{#1}{\protect\numberline{\csname
  the#1\endcsname}{\ignorespaces #2}}\par
  \begingroup
    \@parboxrestore
    \normalsize
    \csname @make#1caption\endcsname
    {\csname fnum@#1\endcsname}{\ignorespaces #3}\par
  \endgroup}

\long\def\@makefigurecaption#1#2{
 \vskip 10pt                     
 {\small                         
 \setbox\@tempboxa\hbox{\small{\bf#1: }#2}
 \ifdim \wd\@tempboxa >\hsize
   \unhbox\@tempboxa\par
 \else
   \hbox to\hsize{\hfil\box\@tempboxa\hfil}       
 \fi}
 \vskip 10pt}                    

\long\def\@maketablecaption#1#2{
 \vskip 10pt                     
 {\small                         
 \setbox\@tempboxa\hbox{\small{\bf#1: }#2}
 \ifdim \wd\@tempboxa >\hsize
   \unhbox\@tempboxa\par
 \else \hbox  to\hsize{\box\@tempboxa\hfil}   
 \fi}
 \vskip 10pt}                    

\makeatother

\def\vec#1{\ifmmode
\mathchoice{\mbox{\boldmath$\displaystyle\bf#1$}}
{\mbox{\boldmath$\textstyle\bf#1$}}
{\mbox{\boldmath$\scriptstyle\bf#1$}}
{\mbox{\boldmath$\scriptscriptstyle\bf#1$}}\else
{\mbox{\boldmath$\bf#1$}}\fi}

\def\math#1{\ifmmode
\mathchoice{\mbox{$\displaystyle\rm#1$}}
{\mbox{$\textstyle\rm#1$}}
{\mbox{$\scriptstyle\rm#1$}}
{\mbox{$\scriptscriptstyle\rm#1$}}\else
{\mbox{$\rm#1$}}\fi}

\def\parth#1{\left(#1\right)}			
\def\brak#1{\left[#1\right]}			
\def\set#1{\left\{#1\right\}}			
\def\avrg#1{<#1>}				
\def\pd#1{\partial_{#1}}			
\def\evaluate#1#2{\left.#1\right|_{#2}}		

\def\accol2#1#2{\left\{ 
\begin{array}{l}#1\\#2\end{array}
\right.}					
\def\matrix22#1#2#3#4{\left(
\begin{array}{cc}#1&#2\\#3&#4\end{array}
\right)}					

\def\pscal#1#2{\left(#1\vert#2\right)}		
\def\abs#1{\left|#1\right|}			
\def\dpart#1#2{\frac{\partial#1}{\partial#2}}	
\def\transp#1{#1^{\math{T}}}			
\def\Order#1{O(#1)}				
\def\order#1{o(#1)}				
\def\limit#1#2{\lim_{#1\rightarrow #2}}		
\def\icx{\math{i}}				
\def\e{\math{e}}				
\def\Re{\math{Re}}				
\def\Im{\math{Im}}				

\def\Tr{\math{Tr}}				
\def\Asin{\math{Arcsin}}			
\def\cotg{\math{cotg}}				

\def\bbbr{{\rm I\!R}} 				

\def\bbbone{{\mathchoice {\rm 1\mskip-4mu l} {\rm 1\mskip-4mu l}
{\rm 1\mskip-4.5mu l} {\rm 1\mskip-5mu l}}}	

\def\bbbq{{\mathchoice {\setbox0=\hbox{$\displaystyle\rm Q$}\hbox{\raise
0.15\ht0\hbox to0pt{\kern0.4\wd0\vrule height0.8\ht0\hss}\box0}}
{\setbox0=\hbox{$\textstyle\rm Q$}\hbox{\raise
0.15\ht0\hbox to0pt{\kern0.4\wd0\vrule height0.8\ht0\hss}\box0}}
{\setbox0=\hbox{$\scriptstyle\rm Q$}\hbox{\raise
0.15\ht0\hbox to0pt{\kern0.4\wd0\vrule height0.7\ht0\hss}\box0}}
{\setbox0=\hbox{$\scriptscriptstyle\rm Q$}\hbox{\raise
0.15\ht0\hbox to0pt{\kern0.4\wd0\vrule height0.7\ht0\hss}\box0}}}}
\def\sq{\hbox{\rlap{$\sqcap$}$\sqcup$}}
\def\qed{\ifmmode\sq\else{\unskip\nobreak\hfil
\penalty50\hskip1em\null\nobreak\hfil\sq
\parfillskip=0pt\finalhyphendemerits=0\endgraf}\fi}	

\def\dQ{\partial Q}
\def\Lb{\abs{\dQ}}
\def\Larc{{\cal L}}
\def\half{\frac{1}{2}}
\def\eps{\varepsilon}
\def\rmin{\rho_{\math{min}}}
\def\rmax{\rho_{\math{max}}}
\def\ds{\displaystyle}
\def\Iw{I_\omega}
\def\C{{\cal C}}
\def\D{{\cal D}}
\def\Ebar{\widehat{E}}
\def\Etilde{\widetilde{E}}
\def\sbar{\widehat{s}}
\def\stilde{\widetilde{s}}
\def\phb{\widehat{\varphi}}
\def\pht{\widetilde{\varphi}}
\def\th{\theta}
\def\ph{\varphi}
\def\lenarc#1{\overline{#1}}
\def\propsep{\vspace{3mm}}
\def\pim0{(\ph_0,I_0,\mu)}
\def\Pim1{(\ph_1,I_1,\mu)}
\def\pjm0{(\psi_0,J_0,\mu)}
\def\im0{(I_0,\mu)}
\def\jm0{(J_0,\mu)}
\def\W0{\Omega_0}
\def\ThNbar{\overline{\Theta}_N}

\begin{document}

\title{Integrability and Ergodicity of Classical Billiards in a Magnetic
Field}

\author{
N. Berglund, H. Kunz \\
{\it Institut de Physique Th\'eorique} \\
{\it Ecole Polytechnique F\'ed\'erale de Lausanne} \\
{\it PHB-Ecublens, CH-1015 Lausanne, Switzerland} \\
{\rm e-mail: }{\tt berglund@iptsg.epfl.ch, kunz@eldp.epfl.ch} \\
}

\date{
Revised version May 1995\footnote{Submitted to J. Stat. Phys. 
December 1994, revised version accepted May 1995,
to be published in J. Stat. Phys., Vol. 83 1/2, April 1996.}
}

\maketitle

\begin{abstract}
We consider classical billiards in plane, connected, but not necessarily
bounded domains. The charged billiard ball is immersed in a homogeneous, 
stationary magnetic field perpendicular to the plane.

The part of dynamics which is not trivially integrable can be described by 
a ``bouncing map''. We compute a general expression for the Jacobian 
matrix of this map, which allows to determine stability and bifurcation 
values of specific periodic orbits.
In some cases, the bouncing map is a twist map and admits a generating
function. We give a general form for this function which is useful to do 
perturbative calculations and to classify periodic orbits.

We prove that billiards in convex domains with sufficiently smooth 
boundaries possess invariant tori corresponding to skipping trajectories.
Moreover, in strong field we construct adiabatic invariants over 
exponentially large times.
To some extent, these results remain true for a class of non-convex
billiards.

On the other hand, we present evidence that the billiard in a square is 
ergodic for some large enough values of the magnetic field. A numerical 
study reveals that the scattering on two circles is essentially chaotic. 
\end{abstract}

\noindent{\large {\bf Key words:} billiards, magnetic field, twist map, 
 integrability, \\ adiabatic invariant, ergodicity}


\section{Introduction}
\label{sec_intr}


We consider the classical motion of a particle of mass $m$ and charge $q$ 
in a plane domain $Q$. A homogeneous, stationary magnetic field $B$ 
perpendicular to the plane makes the particle move on arcs of Larmor 
radius $\mu$. Whenever it encounters the boundary, the particle is  
reflected specularly. This problem was first considered by Robnik and Berry 
(\cite{BeRo}, \cite{Ro}).

One point of interest in such models is the problem of integrable versus 
ergodic behaviour of Hamiltonian systems. In zero field, we know examples of 
integrable billiards (elliptic or rectangular boundary) as well as ergodic 
ones (dispersive Sinai billiards, Bunimovich stadium, see for instance 
\cite{KT}). Since it is unlikely 
that integrability is stable with respect to perturbation by a magnetic field, 
$B$ is a natural parameter for studying the transition from order to chaos.
One question is whether the billiard can become globally ergodic in strong 
enough field.

Another motivation for studying magnetic billiards is connected to the 
problem of quantum chaos. Recently, it has become possible to create 
mesoscopic systems where the electrons' motion is essentially ballistic 
\cite{Levy}. Behaviour of some macroscopic observables, like the 
susceptibility, can be surprisingly complicated \cite{Reza}, and may be 
related to classical dynamics. In order to be able to apply semi-classical 
methods like trace formulae, it is desirable to have a 
good knowledge of classical periodic orbits and their stability, at least 
perturbatively.

This paper is organized as follows. In section \ref{sec_def}, we give a
more precise definition of the billiard flow and boundaries considered.
The interesting part of dynamics can be described by a {\em bouncing map} 
$T$, defined in section \ref{sec_map}, where we also give an exact 
expression for the Jacobian matrix of $T$. In some cases, $T$ is a {\em twist 
map} and admits a {\em generating function} $G$. In section \ref{sec_gen}, we 
give an exact expression of $G$, and we discuss its physical interpretation 
and its practical applications. 

Our main analytical results are contained in section \ref{sec_convex}, 
where we consider billiards with smooth, convex boundaries.
Using perturbative techniques, we analyze different quasi-integrable limits.
In section \ref{sec_nearbound}, we prove the existence of invariant curves 
for all values of the magnetic field. These quasi-periodic trajectories 
correspond physically to diamagnetic currents along the boundary. 
Our KAM type result generalizes the well known theorem of Lazutkin
\cite{Lazutkin} on the existence of caustics near the boundary, in the 
zero field case. The magnetic field however breaks the 
symmetry between the forward and backward skipping orbits, and
at intermediate values of the field, only one kind of skipping 
orbit is present.
In section \ref{sec_strongfield}, we analyze the strong magnetic field
limit. We compute an expansion of the bouncing map in powers of the Larmor 
radius and construct an adiabatic invariant on a time scale growing 
exponentially with the magnetic field when the boundary is analytic. 
At low order, this adiabatic invariant coincides with the one derived by 
Robnik and Berry. 
In order to construct the invariant, we derive a theorem on a class of maps 
of the annulus, which may be of interest in a broader context. 
Section \ref{sec_theo} is devoted to the proof of it. 

In section \ref{sec_nonsmooth}, we study billiards which may show an 
important chaotic component, because of singularities in the boundary 
(polygons) or because of a negative curvature. It turns out that the 
properties of billiards in convex domains can be to some extent 
generalized to billiards with concave boundary. In general it seems 
that billiards in a magnetic field are of the mixed type (invariant 
tori and chaotic components coexist). A possible exception is the 
case of the square at some particular values of the magnetic field, 
where we present an analytical argument and numerical evidence for a 
completely chaotic motion.


\section{Definition of the billiards}
\label{sec_def}


Let $Q$ be a connected domain in $\bbbr^2$. We assume for simplicity that 
the boundary $\dQ$ consists of simple, closed, piecewise $C^2$ curves of 
total length $\Lb$, although some of our results may be extended to more 
general cases. Several of them however need a higher degree of 
differentiability, which will be clearly indicated on the spot.

\begin{figure}
 \centerline{\psfig{figure=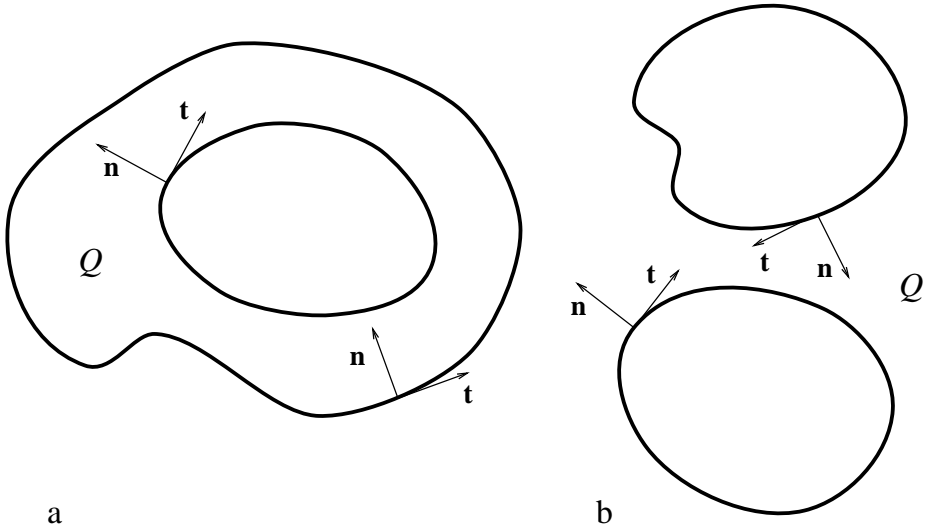,height=80mm}}
 \caption{Examples of billiard domains $Q$, with tangent and normal vectors to 
	 the boundary $\dQ$: (a) bounded domain, (b) unbounded domain.}
\label{fig_bords}
\end{figure}

The boundary is parameterized with the {\em curvilinear abscissa} or 
{\em arclength} $s$:
\begin{equation}
 \vec{x}(s) = (X(s), Y(s)), \;\; ds^2 = dX^2 + dY^2, \;\; s \in [0, \Lb).
 \label{def1}
\end{equation}
The unit tangent and normal vectors and the (signed) curvature are given by
\begin{eqnarray}
 \vec{t}(s)&=&(X'(s), Y'(s)) = (\cos\tau(s), \sin\tau(s)), \nonumber \\
 \vec{n}(s)&=&(-Y'(s), X'(s)),  \nonumber \\
 \kappa(s)&=&\frac{d\tau}{ds} = X'(s)Y''(s)- X''(s)Y'(s) = \frac{1}{\rho(s)}.
 \label{def2} 
\end{eqnarray}
Parameterization is chosen in such a way that $\vec{n}$ is always oriented
towards the interior of $Q$ (see fig.\ref{fig_bords}). 
In this way, the curvature is positive for a convex boundary. We suppose 
$\kappa(s)$ to be defined and continuous everywhere but on a set $E_2$ of 
punctual values of $s$. The vectors $\vec{t}$ and $\vec{n}$ are undefined on 
a set $E_1 \subset E_2$.

Inside $Q$, the billiard flow is given by the Lagrangian
\begin{equation}
 L(\vec{x}, \dot{\vec{x}}) = \half m\vec{\dot{x}}^2 + 
 q\pscal{\dot{\vec{x}}}{\vec{A}(\vec{x})}, \;\;
 \vec{A}(\vec{x}) = (-\half y B,\half x B).
 \label{def3}
\end{equation}
The resulting motion is simply circular uniform, with Larmor radius
\begin{equation}
 \mu = \frac{m v}{\abs{q B}} = \frac{\sqrt{2 m E}}{\abs{q B}},
 \label{def4}
\end{equation}
where speed $v=\abs{\dot{\vec{x}}}$ and energy $E$ are constants of motion.
We adopt the sign convention $qB < 0$, which implies that the particle turns 
counterclockwise, and
we take some characteristic length of the billiard as a length unit, so that 
$\mu$ can be considered as a dimensionless parameter. Thus, increasing 
$B$ is equivalent to decreasing $E$ or increasing the size of the billiard
(which can be viewed as taking the thermodynamic limit).

Each time the particle hits the boundary, the velocity changes according to 
the law of specular reflection 
$\vec{v}' = \vec{v} - 2 \pscal{\vec{v}}{\vec{n}} \vec{n}$, which is well 
defined for almost every point of the boundary\footnote{Subtleties may occur 
in the corners or if $\pscal{\vec{v}}{\vec{n}}=0$ and $\rho\leq\mu$ at the point 
of collision, but they are of little importance for the following.}.

The phase space $M = Q\times S^1$ can be divided into two disjoint sets 
$M_1$ and $M_2$. $M_1$ consists of all the orbits that never touch the boundary.
It corresponds to an integrable component of the motion and may be empty.
In non-zero magnetic field ($\mu < \infty$), the orbits of $M_2$ hit the 
boundary an infinite number of times. It is thus natural to study the 
dynamics in $M_2$ by means of a bouncing map.


\section{The bouncing map and its Jacobian matrix}
\label{sec_map}

\begin{figure}
 \centerline{\psfig{figure=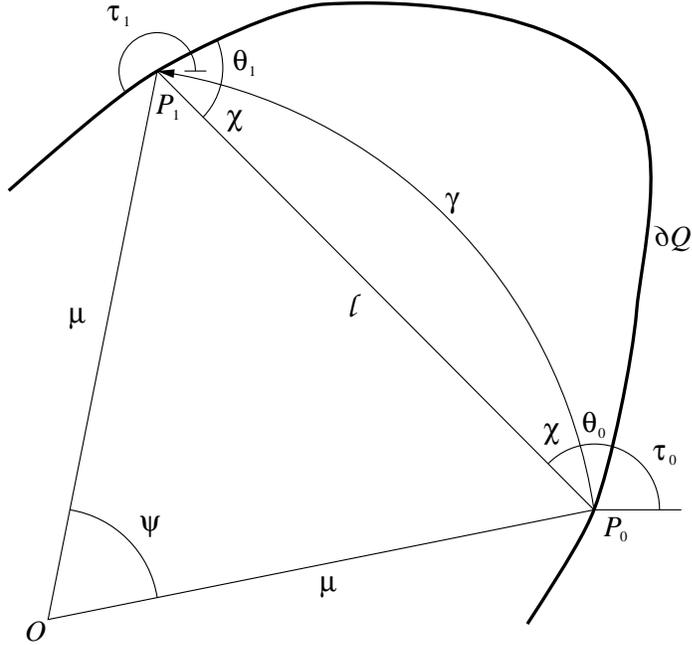,height=90mm}}
 \caption{The trajectory between two successive bounces is an arc $\gamma$ of 
	 radius $\mu$, angle $\psi$ and extremities $P_0$ and $P_1$.
	 The Jacobian matrix of the bouncing map $T$ depends on the length 
	 $\ell$ of the chord, the angle $\chi$ between chord and arc,
	 the angles $\theta_0$, $\theta_1$ between arc and boundary 
	 and the curvature at $P_0$, $P_1$.}
\label{fig_arc}
\end{figure}

Let us consider the trajectory between two successive collisions with $\dQ$, 
occurring at $P_0 = (X(s_0), Y(s_0))$ and $P_1 = (X(s_1), Y(s_1))$. 
The trajectory is an arc $\gamma$ of center $O$, radius $\mu$ and angle $\psi$ 
(see fig.\ref{fig_arc}).

We call $\theta_i$ the angle between the arc and the boundary at $P_i$, 
and $u_i = -\cos{\theta_i}$, $i = 0,1$. Quantities $s$ and $u$ are the 
{\em Birkhoff variables} and the {\em bouncing map} $T$ is defined as
\begin{equation}
 T: (s_0, u_0) \mapsto (s_1, u_1).
 \label{map1}
\end{equation}

If $\ell$ is the length of the chord $P_0P_1$, and $\chi$ is the angle 
between the chord and the arc $\gamma$, simple geometry shows that
\begin{equation}
 \psi = 2\chi, \;\; \sin\chi = \frac{\ell}{2\mu}.
\label{map1b}
\end{equation}
In general, there may be {\em two} trajectories with supplementary $\chi$ 
for a given $\ell$ (see also fig.\ref{fig_strfield}). 
This is a characteristic magnetic field effect.

Generalizing Birkhoff's technique (\cite{Birk}, p.173, see also 
\cite{KT}), we find

\propsep

\noindent 
{\bf Proposition 1}
{\it If $s_0, s_1 \not\in E_2$, the Jacobian matrix $DT$ of $T$ is given by
\begin{eqnarray}
 \label{map2}
 \dpart{s_1}{s_0}&=&
 \frac{\kappa_0 \ell \cos\chi - \sin(\theta_0 + 2\chi)}{\sin\theta_1} \nonumber \\
 \dpart{s_1}{u_0}&=&
 \frac{\ell \cos\chi}{\sin\theta_1 \sin\theta_0} \nonumber \\ 
 \dpart{u_1}{s_0}&=&
 \frac{\sin(\theta_0 + 2\chi)\sin(\theta_1 + 2\chi) - \sin\theta_0 \sin\theta_1}{\ell\cos\chi}
 \nonumber \\ 
 && - \kappa_0\sin(\theta_1 + 2\chi) -  \kappa_1\sin(\theta_0 + 2\chi) + \kappa_0\kappa_1\ell\cos\chi
 \nonumber \\ 
 \dpart{u_1}{u_0}&=&
 \frac{\kappa_1 \ell \cos\chi - \sin(\theta_1 + 2\chi)}{\sin\theta_0}
\end{eqnarray}
where $\kappa_i = \kappa(s_i)$.} 

\propsep

For completeness and consistency of notation, we give a proof in appendix 
\ref{app1}. We noticed that an alternative proof 
of this formula was provided by \cite{MBG}.
All the quantities appearing in (\ref{map2}) can be easily expressed as 
functions of $s_0$ and $s_1$ (see (\ref{A1}) and (\ref{A2})). 
It is in general much more difficult to solve 
the relation $u_0(s_0, s_1)$ with respect to $s_1$, in order to obtain $DT$ as 
a function of $z_0 = (s_0, u_0)$.

A first observation is that $DT$ has unit determinant, and hence the Birkhoff 
variables are conjugate, although momentum and velocity are not collinear in 
a magnetic field (see section \ref{sec_gen}).

A second observation concerns the element $\dpart{s_1}{u_0}$. We see that 
its sign depends only on $\chi$. Suppose that the shape of $Q$ is such 
that it cannot contain any arc of radius $\mu$ and angle larger than $\pi$. 
Then we have $\dpart{s_1}{u_0} > 0$, and $u_0$ is uniquely defined for 
given $s_0$ and $s_1$. In such a case, $T$ is called a (symplectic) 
{\em twist map}. Twist maps, especially when continuous, have a lot of 
properties (see e.g. \cite{Meiss} for a review), some of which we 
will discuss in the next section.

The Jacobian matrix allows us to compute Liapunov exponents and to determine 
stability of periodic orbits. If we define for $z = (s, u)$ the {\em stability 
matrix}
\begin{equation}
 S_n(z) = DT(T^{n-1}z)\cdots DT(Tz)DT(z),
\label{map3}
\end{equation}
then the Liapunov exponents are given by
\begin{equation}
 \Lambda_\pm(z) = \pm\lim_{n\rightarrow\infty} \frac{1}{2n}\ln\Tr \transp{S_n}(z)S_n(z).
\label{map4}
\end{equation}

If $z$ belongs to an orbit of period $n$, the orbit is hyperbolic and unstable 
if $\abs{\Tr S_n(z)}>2$, parabolic if $\abs{\Tr S_n(z)}=2$ and elliptic if 
$\abs{\Tr S_n(z)}<2$. In the latter case, the orbit is stable unless resonance 
occurs. In section \ref{sec_nonsmooth}, we will give examples of how to apply 
the formula (\ref{map2}) to analyze stability of specific families of 
periodic orbits.


\section{Generating functions}
\label{sec_gen}


Here we call {\em generating function} of the continuous map $T$ a $C^1$ 
function $G(s_0, s_1)$ such that
\begin{equation}
 dG = u_0 ds_0 - u_1 ds_1.
\label{gen1}
\end{equation}

It is easy to check the following properties \cite{Meiss}:
\begin{enumerate}
 \item	If $T$ is an area preserving twist map, it admits a generating function, 
	unique up to an additive constant, given by
\begin{equation}
 G(s_0, s_1) = \int^{(s_0, s_1)} u_0(\xi, \eta)d\xi - u_1(\xi, \eta)d\eta.
\label{gen2}
\end{equation}

 \item 	If $G$ is $C^2$, the map $T$ generated by $G$ is always area preserving.
	It is a twist map if $\partial^2_{s_0s_1}G > 0$ \footnote{$T$ can 
	be degenerate if $\partial^2_{s_0s_1}G = 0$.}.

 \item 	If $G$ is $C^2$, $u$ is a constant of motion iff $G(s_0, s_1) = g(s_1-s_0)$.
\end{enumerate}

In zero field, $G$ is known to be the length of the chord $\ell(s_0, s_1)$. 
For magnetic billiards, we found that $G$ depends also on an area associated 
to the trajectory, a feature appearing apparently in all problems 
involving a magnetic field:

\propsep

\noindent 
{\bf Proposition 2}
{\it Suppose that $Q$ is bounded and that $T$ is a twist map. Then, for 
$s_0, s_1 \not\in E_2$, the generating function is given by 
\begin{equation}
 G = \Larc + \frac{1}{\mu} {\cal S},
\label{gen3}
\end{equation}
where $\Larc$ is the length of the arc and $\cal S$ is the area between 
the arc and the boundary (see fig.\ref{fig_gen}).}

\propsep

We give the proof in appendix \ref{app2}.
For practical purposes, it is useful to write $G$ in the form
\begin{equation}
 G(s_0, s_1) = \frac{1}{\mu} A(s_0, s_1) + B_{\mu}(\ell(s_0, s_1)),
\label{gen4}
\end{equation}
where $A$, the area between the chord and the boundary, does not depend on the 
magnetic field, and
\begin{equation}
 B_{\mu}(\ell) = \mu\Asin \parth{\frac{\ell}{2 \mu}} 
 + \frac{\ell}{2}\sqrt{1-\frac{\ell^2}{4 \mu^2}}
\label{gen5}
\end{equation}
is equal to $\Larc$ minus the area between the chord and the arc, 
divided by $\mu$.

As an example, let us consider an elliptic boundary, of equation
\begin{equation}
 \vec{x} = (\lambda\cos\varphi, \sin\varphi), \;\;\;
 \frac{ds}{d\varphi} = C(\varphi) = \sqrt{\cos^2\varphi+\lambda^2\sin^2\varphi}.
\label{gen6}
\end{equation}
If the magnetic field is low enough (in fact, if $\mu>\lambda^2$, as we shall 
see in section \ref{sec_convex}), $T$ is a continuous twist map, and
\begin{equation}
 G = \frac{\lambda}{\mu}\brak{\varphi_- - \half\sin(2\varphi_-)}
 + B_\mu(2\sin\varphi_- C(\varphi_+)),
\label{gen7}
\end{equation}
 where $\varphi_\pm = \half(\varphi_1 \pm \varphi_0)$.
Note that for a circular boundary ($\lambda = 1$), $C = 1$ and $G$ depends 
only on $\varphi_1 - \varphi_0$. Thus, the billiard becomes integrable, 
which is geometrically obvious.

\begin{figure}
 \centerline{\psfig{figure=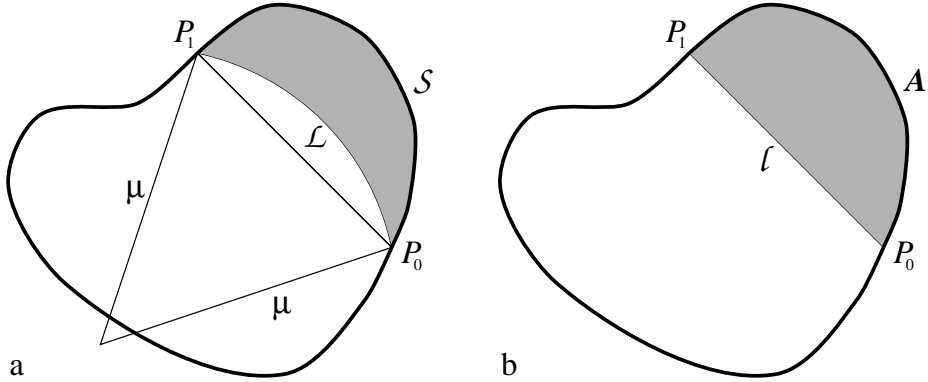,height=50mm}}
 \caption{When $T$ is a twist map, its generating function $G$ can be expressed 
	 (a) as a function of the length $\Larc$ of $\gamma$ and the area 
	 ${\cal S}$ between $\gamma$ and $\dQ$, or 
	 (b) as a function of the 
	 length $\ell$ of $P_0P_1$ and the area $A$ between $P_0P_1$ and $\dQ$.}
\label{fig_gen}
\end{figure}

In order to give a physical interpretation to $G$, let us recall that the 
momentum, which is canonically conjugate to the position, is given by
\begin{equation}
 \vec{p} = \dpart{L}{\vec{\dot{x}}} = m\dot{\vec{x}} + q\vec{A} =
 (m\dot{x} - \half q B y, m\dot{y} + \half q B x).
\label{gen8}
\end{equation} 
Thus, it would have been physically more natural to use, instead of 
the tangent velocity $u$, the tangent momentum
\begin{equation}
 p = -\frac{1}{mv} \pscal{\vec{p}}{\vec{t}(s)} =
 u + \frac{1}{2\mu} (X(s)Y'(s) - X'(s)Y(s)). 
\label{gen9}
\end{equation} 
Indeed, $p$ and $s$ are conjugate since they admit as a generating function 
the reduced action along the arc $\gamma$:
\begin{equation}
 F = \int_\gamma \pscal{\vec{p}}{d\vec{x}}
   =  mv \int_\gamma ds+\frac{1}{2\mu}\parth{ydx-xdy}
 \; \Rightarrow \; dF = mv(p_0 ds_0 - p_1 ds_1).
\label{gen10}
\end{equation} 
It is however more convenient to use $u$ instead of $p$. Green's theorem 
implies that the generating functions are related by 
\begin{equation}
 mvG =  F + \half q B \int_{\gamma_Q} ydx-xdy,
\label{gen10b}
\end{equation}  
where $\gamma_Q$ is the piece of boundary connecting $P_0$ to $P_1$.

One application of generating functions is in perturbation theory.
Suppose the map $T_\eps$ depends on a parameter $\eps$ (controlling the 
magnetic field or the shape of the boundary), such that the behaviour of $T_0$ 
is known (e.g. integrable). Approximating $T_\eps$ by 
its expansion $T_0 + \eps \partial_\eps T_\eps \vert_{\eps = 0}$ does in 
general not lead to an area preserving map, whereas the map generated 
by $G(s_0, s_1, 0) + \eps \partial_\eps G(s_0, s_1, \eps) \vert_{\eps = 0}$ 
is always conservative.

For example, the generating function for an elliptic boundary close to 
a circle, given by (\ref{gen6}) with $\lambda = 1 + \eps$, is
\begin{eqnarray}
 G &=& G_0(\varphi_-) + \eps G_1(\varphi_+, \varphi_-) + \Order{\eps^2}, \nonumber \\
 G_0(\varphi_-) &=& \frac{1}{\mu}\brak{\varphi_-+\half\sin(2\varphi_-)}
 + B_\mu(2\sin\varphi_-), \nonumber \\
 G_1(\varphi_+, \varphi_-) &=& \frac{1}{\mu}\brak{\varphi_-+\half\sin(2\varphi_-)} 
 + 2\sin\varphi_-\sin^2\varphi_+ \sqrt{1 - \frac{\sin^2\varphi_-}{\mu^2}}.
\label{gen11}
\end{eqnarray}

Another use of generating functions is in searching periodic orbits.
If we define the $n$-point generating function
\begin{equation}
 G^{(n)}(s_0, s_1, \ldots, s_{n-1}) 
 = G(s_0, s_1) + G(s_1, s_2) + \cdots + G(s_{n-1}, s_0),
\label{gen12}
\end{equation}
then the law of specular reflection implies that every periodic orbit containing 
no points with $s \in E_2$ is a solution of
\begin{equation}
 \dpart{G^{(n)}}{s_0} = \dpart{G^{(n)}}{s_1} = \ldots = \dpart{G^{(n)}}{s_{n-1}} = 0,
\label{gen13}
\end{equation}
which is a system of $n$ nonlinear algebraic equations of the $n$ variables  
$s_0, \ldots, s_{n-1}$. 

It is convenient to ``lift'' the periodic variable $s$ to the real line.
An $(m, n)$ periodic orbit is then defined by $s_n = s_0 + m\Lb, u_n = u_0$,
and its frequency $\omega = \frac{1}{\Lb}\limit{k}{\infty} \frac{s_k}{k}$
 is equal to $\frac{m}{n}$. The possible frequencies belong to an interval 
$\Iw$, depending on the behaviour of the boundaries $u=\pm 1$ of the phase 
cylinder.
First results on existence of periodic orbits for continuous area-preserving 
twist maps are due to Poincar\'e and Birkhoff. Powerful developments 
were achieved by Aubry and Le Daeron, Mather, MacKay and Meiss, and Katok (see 
\cite{Meiss} and references therein for more details):

\begin{enumerate}
\item 	For every $m, n, \frac{m}{n} \in \Iw$, there is at least one $(m, n)$ 
	periodic orbit which is ``maximizing''. This means that every finite orbit 
	segment $(s_k, \ldots, s_l), l \geq k+2$ is a global maximum of 
	$\sum_{j=k}^{l-1} G(s_j, s_{j+1})$ with respect to variations of 
	$s_{k+1}, \ldots, s_{l-1}$.
	In particular, $(s_0, \ldots, s_{n-1})$ is a global maximum\footnote{Second 
	variations of $G$ can be computed 
	using (\ref{A5}) and (\ref{A6}).} of $G^{(n)}$.
	If the maximum is nondegenerate, the orbit is hyperbolic.
	
\item	For every $m, n, \frac{m}{n} \in \Iw$, there is at least one $(m, n)$ 
	periodic orbit which is ``maximin''. This means that the Hessian matrix 
	of $\sum_{j=k}^{l-1} G(s_j, s_{j+1})$ has one single positive 
	eigenvalue. The 
	orbit is either elliptic or inverse hyperbolic ($\Tr S_n < -2$).
	
\item	Every orbit on a rotational invariant circle is maximizing.
	For every irrational $\omega \in \Iw$, there is a maximizing 
	quasiperiodic orbit of frequency $\omega$. Its closure is either an 
	invariant circle, or an invariant Cantor set. This result is in some 
	sense stronger than KAM theory, since it shows the existence of 
	quasiperiodic orbits for twist maps that are not necessarily nearly 
	integrable. 
\end{enumerate}

To summarize: if the bouncing map is a continuous twist map, its generating 
function provides a useful tool to do perturbative as well as variational
calculations. In particular, we have a lower bound on the number of periodic 
orbits of period $n$ (namely twice as many as there are integers $m$ coprime 
with $n$ such that $\frac{m}{n} \in \Iw$), whose stability can be related 
to the second derivative of the generating function.

As we shall see in the next section, an important class of maps 
satisfying the twist property is given by bouncing maps of billiards with 
smooth, convex boundaries in low magnetic field. Many billiards' 
bouncing maps however are either not twist or discontinuous.
Nevertheless, generalized versions of the generating function can be useful 
tools in some of these cases as well, as we will see in sections 
\ref{sec_strongfield} and \ref{sec_neg}.


\section{Smooth convex boundaries}
\label{sec_convex}


\begin{figure}
 \centerline{\psfig{figure=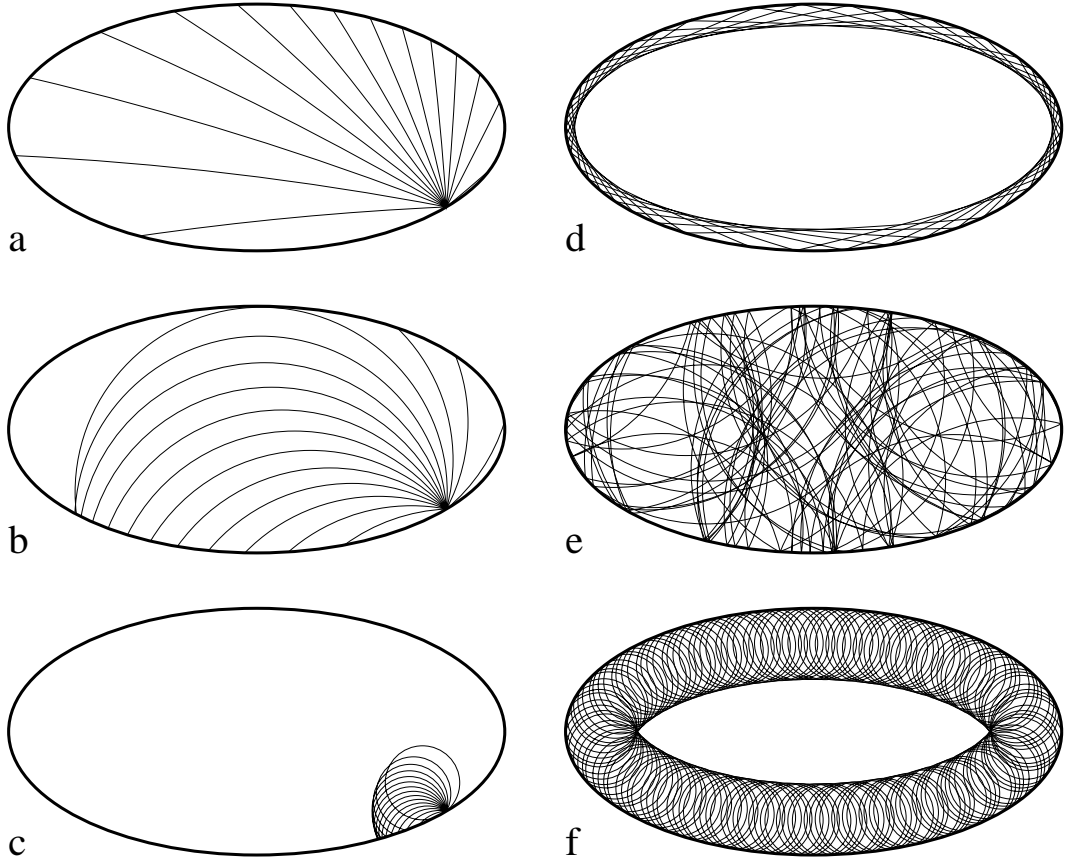,width=140mm}}
 \caption{Trajectories in an elliptic boundary, with constant $s_0$ and 
	 varying $u_0$: 
	 (a)~$\mu \geq \rmax$, 
	 (b)~$\rmin < \mu < \rmax$, 
	 (c)~$\mu \leq \rmin$.
	 Trajectories with $u_0$ near $-1$:
	 (d)~trajectory with caustic for $\mu \geq \rmax$,
	 (e)~chaotic trajectory when $\rmin < \mu < \rmax$,
	 (f)~trajectory with caustic for $\mu \leq \rmin$.}
\label{fig_ellipse}
\end{figure}


\subsection{Qualitative behaviour}
\label{sec_qual}


We say that $\dQ$ is smooth and convex\footnote{For brevity, we omit the
word ``strictly''.} if $\rho(s)$ is a smooth function 
(in a sense that we will have to specify) and is bounded by positive constants:
$0 < \rmin \leq \rho(s) \leq \rmax < \infty \; \forall s$. For example, for the 
ellipse (\ref{gen6}), $\rmin = \lambda^{-1}$ and $\rmax = \lambda^2$.

\begin{figure}
 \centerline{\psfig{figure=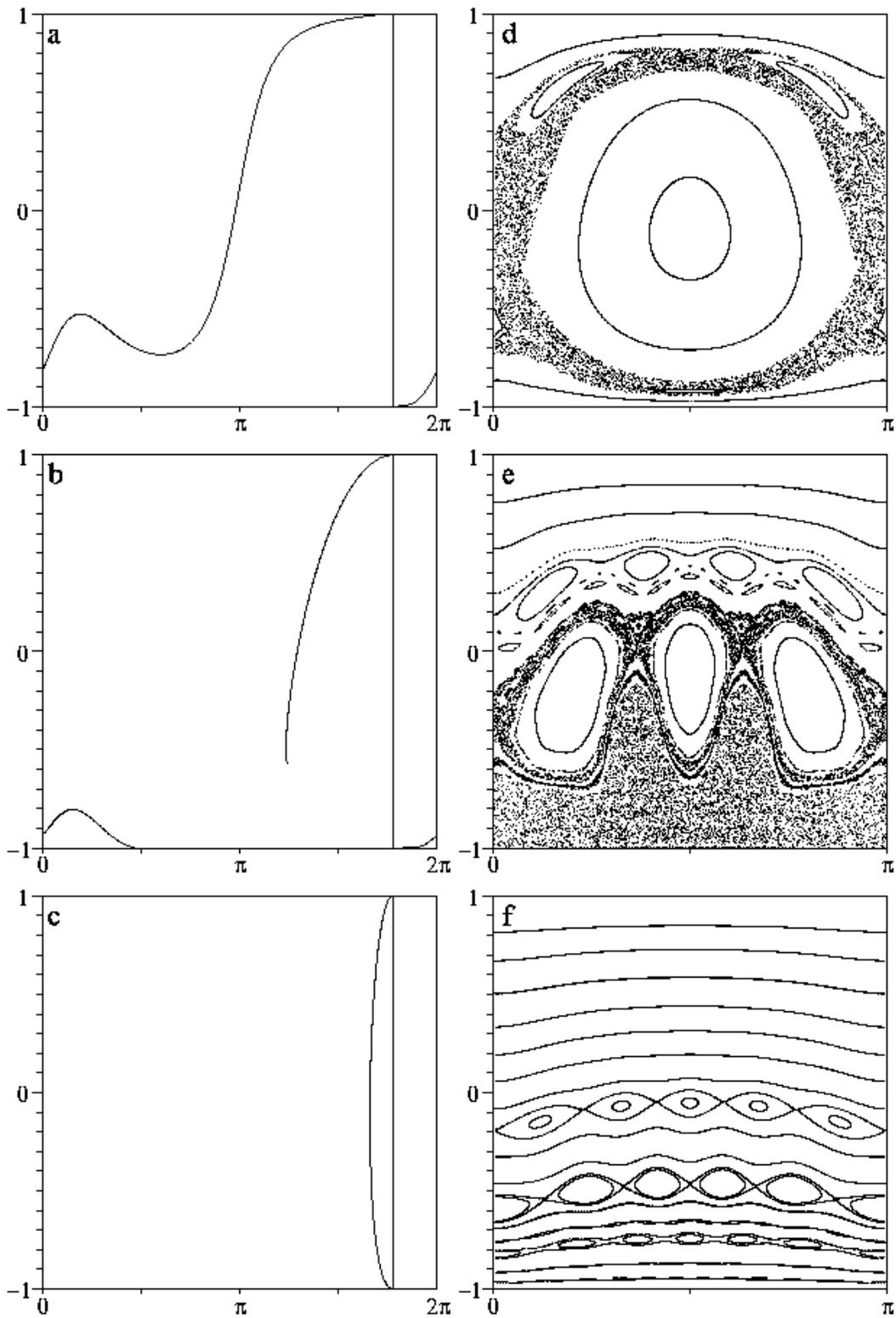,width=140mm,clip=t}}
 \caption{Structure of phase space in the $(\varphi, u)$ plane for the ellipse.
	 Image of the line $(\varphi = \varphi^*, -1 \leq u \leq 1):$ 
	 (a)~$\mu \geq \rmax$, 
	 (b)~$\rmin < \mu < \rmax$, 
	 (c)~$\mu \leq \rmin$.
	 Typical phase portraits:
	 (d)~$\mu \geq \rmax$,
	 (e)~$\rmin < \mu < \rmax$,
	 (f)~$\mu \leq \rmin$.}
\label{fig_mapell}
\end{figure}

Robnik and Berry \cite{BeRo} proposed to classify the dynamics by comparing
$\mu$ to $\rmin$ and $\rmax$. If we consider specifically the twist 
property, by looking at trajectories with fixed $s_0$ and varying $u_0$ 
(fig.\ref{fig_ellipse} and \ref{fig_mapell}), we obtain the following 
``curvature regimes'':

\begin{enumerate}
 \item	If $\mu \geq \rmax$, the function $s_1(s_0, u_0)$ is increasing in $u_0$, 
	so that $T$ is a {\em twist map}. The curve $\set{T(s_0, u_0), 
	-1 < u_0 < 1}$ turns exactly once around the phase cylinder, with 
	$\limit{u_0}{\pm 1} T(s_0, u_0) = (s_0, u_0)$.
	Thus, $\Iw = [0, 1]$.

 \item	If $\rmin < \mu < \rmax$, discontinuities in the map occur when the 
	trajectory becomes tangent to the boundary. We have  
	$\limit{u_0}{+1} T(s_0, u_0) = (s_0, u_0)$, but this is not 
	necessarily true for $u_0 \rightarrow -1$.
	In appendix \ref{app4}, we illustrate the construction of 
	the discontinuity lines of the map.

 \item	If $\mu \leq \rmin$, the function $s_1(s_0, u_0)$ is first decreasing 
	in $u_0$, reaches its minimum when $\chi = \pi/2$, and increases 
	again. We again have $\limit{u_0}{\pm 1} T(s_0, u_0) = (s_0, u_0)$, 
	so that there are exactly two trajectories with supplementary 
	$\chi$ for given $s_0$, $s_1$.
	The map is no longer twist. One can expect that the frequencies 
	belong to an interval $\Iw = [\omega_{\math{min}}, 0]$, where
	$\omega_{\math{min}} \sim \mu$.
\end{enumerate}

Figures \ref{fig_ellipse} and \ref{fig_mapell} illustrate this behaviour 
in the case of an elliptic boundary\footnote{The billiard in an ellipse 
has the particularity to possess two integrable limits. In the circle 
limit $\lambda=1$, $u$ is a constant of motion, as we noted in the previous
section. In the zero field limit $\mu=\infty$, the product of the angular 
momenta with respect to the foci is a constant, which can be written 
$L=C^2(\ph)(u^2-1)+1$.}. We see that there always exist invariant 
curves near $u = +1$, corresponding to trajectories with caustics (the 
``whispering gallery modes''). For $\mu \geq \rmax$ and $\mu \leq \rmin$, 
such curves also exist near $u = -1$. However, when $\rmin < \mu < \rmax$, 
the tori near $u=-1$ are replaced by a chaotic region. 
This behaviour can be heuristically 
understood by noting the following two points: first, discontinuities 
of the map due to tangencies may destroy invariant curves in an analog manner 
as do the discontinuities due to corners of the boundary \cite{DaRo};
second, the Jacobian matrix (\ref{map2}) diverges like $(u+1)^{-1/2}$, 
causing strong dispersion of nearly tangent orbits, which can be a source of 
positive Liapunov exponents.

Robnik and Berry have computed an {\em adiabatic invariant}, which tends to 
be conserved if $\abs{\frac{\kappa \mu \sqrt{1-u^2}}{1+\kappa \mu u}}\ll 1$:
\begin{equation}
 K(s, u) = \frac{\mu \kappa(s) + u(3-2u^2)}{(1-u^2)^{3/2}},
\label{convex1}
\end{equation}
and gives a very good approximation of the invariant curves for 
$u \rightarrow \pm 1$. As it turns out, in the strong field limit 
$\mu \kappa \ll 1$, $K$ tends to be conserved for {\em all} orbits, so that 
this limit seems to be integrable. In the following, our aim is to 
delineate the validity of these statements and make them
more precise by using different perturbative approaches.

 
\subsection{Near the boundary}
\label{sec_nearbound}


We recall that for Euclidean billiards, the existence of 
invariant curves and caustics near the boundary was first 
demonstrated by Lazutkin \cite{Lazutkin}, who had to assume a high 
degree of differentiability. Douady \cite{Douady} reduced the 
required degree to 6. On the other hand, Hubacher \cite{Hubacher} 
proved that no caustics exist near a boundary whose curvature 
is discontinuous, and Mather \cite{Mather} found that if the curvature
of the boundary vanishes at one point, then invariant curves are 
totally absent.

We are now going to investigate the behaviour of the bouncing map 
of billiards in non-zero magnetic field near their boundary, i.e. for 
small $\sin \th_0$. In this way, we will be able to apply KAM theorems 
to show the existence of invariant curves. The results are 
summarized in theorem 1 at the end of this section.

If we write $z(s)=X(s)+\icx Y(s)$, and $z_i=z(s_i)$, 
$\tau_i=\tau(s_i)$, we have
\begin{equation}
z_1-z_0 = \int_{s_0}^{s_1} \e^{\icx\tau(s)} ds 
= \int_{\tau_0}^{\tau_1} \rho(\tau) \e^{\icx\tau} d\tau.
\label{nearb1}
\end{equation}

Since the $z_i$ also belong to the arc of trajectory, which
has tangent directions $\tau_0+\th_0$ at $z_0$ and 
$\tau_1-\th_1$ at $z_1$ (see fig.\ref{fig_arc}), we have
\begin{equation}
z_1-z_0 = \frac{\mu}{\icx}\brak{\e^{\icx(\tau_1-\th_1)}-
\e^{\icx(\tau_0+\th_0)}}.
\label{nearb2}
\end{equation}
Introducing $A=\e^{-\icx\tau_0}(z_1-z_0)$ and 
$\delta=\tau_1-\tau_0-\th_1+\th_0$, this can be rewritten
\begin{equation}
A - \mu\e^{-\icx\th_0}\frac{\e^{\icx\delta}-1}{\icx}
= -2\mu\sin\th_0,
\label{nearb3}
\end{equation}
which is equivalent to the system
\begin{eqnarray}
\frac{\Re A}{\sin\th_0} - \mu\cos\th_0\frac{\sin\delta}{\sin\th_0}
+\mu(\cos\delta-1) & = & -2\mu \nonumber \\
\frac{\Im A}{\sin^2\th_0} + \mu\frac{\sin\delta}{\sin\th_0}
+\mu\cos\th_0\frac{\cos\delta-1}{\sin^2\th_0} & = & 0.
\label{nearb4}
\end{eqnarray}
If the boundary is $C^k$, this is a system of $C^{k-1}$ 
equations in the variables $s_0,\th_0,s_1,\th_1$ that we 
would like to solve with respect to $s_1,\th_1$.

Writing $s_1=s_0+\sin\th_0\mu\sigma$ and 
$\th_1=\th_0+\sin\th_0\mu\gamma$, we obtain for $k\geq 3$
\begin{eqnarray}
A & = & \sin\th_0\mu\sigma + \frac{\icx}{2}
\sin^2\th_0\mu^2\sigma^2\kappa(s_0) +
\Order{\mu^3\sin^3\th_0} \nonumber \\ 
\delta & = & \sin\th_0\mu\brak{\kappa(s_0)\sigma-\gamma} + 
\Order{\mu^2\sin^2\th_0}. 
\label{nearb5}
\end{eqnarray}
Replacing this in (\ref{nearb4}) finally leads to the system
\begin{eqnarray}
\sigma - \mu\cos\th_0\brak{\kappa(s_0)\sigma-\gamma}
+\Order{\sin\th_0} & = & -2 \nonumber \\
\sigma^2\kappa(s_0) + 2\brak{\kappa(s_0)\sigma-\gamma}
-\mu\cos\th_0\brak{\kappa(s_0)\sigma-\gamma}^2
+\Order{\sin\th_0} & = & 0,
\label{nearb6}
\end{eqnarray}
which has the solution
\begin{eqnarray}
\sigma(s_0,\th_0,\mu) & = & 
\frac{-2}{1-\mu\cos\th_0\kappa(s_0)} + \Order{\sin\th_0} 
\nonumber \\
\gamma(s_0,\th_0,\mu) & = & \Order{\sin\th_0}.
\label{nearb7}
\end{eqnarray}
Since the Jacobian of (\ref{nearb6}) evaluated at this 
solution is $-2 + \Order{\sin\th_0}$, the implicit function 
theorem implies that for small $\sin\th_0$, the bouncing 
map is $C^{k-1}$ and has the form
\begin{eqnarray}
s_1 & = & s_0
-\frac{2\mu\sin\th_0}{1-\mu\cos\th_0\kappa(s_0)} +
\order{\sin\th_0} \pmod{\Lb} \nonumber \\
\th_1 & = & \th_0 + \order{\sin\th_0}.
\label{nearb8}
\end{eqnarray}
However, this formula is only valid if we are able to check 
two properties: firstly, the approximation must be well defined, 
i.e. the denominator in the first equation should not vanish. 
Secondly, the solution (\ref{nearb7}) has to be the ``right'' one, 
that is the {\em first} intersection of the trajectory with the boundary,
following the motion of the particle. We have to distinguish 
between the following cases:
\begin{enumerate}
\item {\bf Near $u=+1$:} Writing $\th=\pi-\eta$, 
(\ref{nearb8}) becomes
\begin{eqnarray}
s_1 & = & s_0-\frac{2\mu}{1+\mu\kappa(s_0)}\eta_0 +
\order{\eta_0} \pmod{\Lb} \nonumber \\
\eta_1 & = & \eta_0 + \order{\eta_0}.
\label{nearb9}
\end{eqnarray}
Here the denominator can never vanish. Moreover, this solution 
corresponds geometrically to a short skip of the particle 
backwards along the boundary, so that it certainly describes 
the first encounter of the trajectory with $\dQ$.

Note that this equation is well defined even when the 
curvature is allowed to vanish. Only when $\mu=\infty$, 
the first equation becomes $s_1=s_0-\frac{2}{\kappa(s_0)}\eta_0
+\order{\eta_0}$, and diverges when $\kappa\rightarrow 0$, 
which is coherent with Mather's result.

The change of variables $\ph=s+\mu\tau(s)$, $r=2\mu\eta$ 
is well defined and transforms the map into
\begin{eqnarray}
\ph_1 & = & \ph_0-r_0 + \order{r_0} 
\pmod{\Lb+2\pi\mu} \nonumber \\
r_1 & = & r_0 + \order{r_0}.
\label{nearb10}
\end{eqnarray}

\item {\bf Near $u=-1$:} The map has the form
\begin{eqnarray}
s_1 & = & s_0-\frac{2\mu}{1-\mu\kappa(s_0)}\th_0 +
\order{\th_0} \pmod{\Lb} \nonumber \\
\th_1 & = & \th_0 + \order{\th_0}.
\label{nearb11}
\end{eqnarray}
This time, we have to be careful with the denominator, 
which may vanish. The approximation $\th\ll 1$ is only 
valid in two cases:
\begin{enumerate}
\item If $\mu\geq\rmax(1+\eps)$, $\eps>0$, then in the variables 
$\ph = \mu\tau(s)-s$, $r=2\mu\th$, the map becomes
\begin{eqnarray}
\ph_1 & = & \ph_0+r_0 + \order{r_0} 
\pmod{2\pi\mu-\Lb} \nonumber \\
r_1 & = & r_0 + \order{r_0}.
\label{nearb12}
\end{eqnarray}
We again got the right intersection, because the particle 
is skipping forward along the boundary (fig.\ref{fig_ellipse}d).

\item If $\mu\leq\rmin(1-\eps)$, then in the variables 
$\ph = s-\mu\tau(s)$, $r=2\mu\th$, the map reads
\begin{eqnarray}
\ph_1 & = & \ph_0-r_0 + \order{r_0} 
\pmod{\Lb-2\pi\mu} \nonumber \\
r_1 & = & r_0 + \order{r_0}.
\label{nearb13}
\end{eqnarray}
This time, the particle is starting with a forward glancing velocity, 
but reaches the boundary behind its starting point
(fig.\ref{fig_ellipse}f). It is not obvious 
that there is not another intersection of the trajectory and the boundary 
in between. Luckily, lemma 3 in appendix \ref{app5} shows that 
this cannot happen, because any circle of radius $\mu<\rmin$ cuts
$\dQ$ at two points at most.

\end{enumerate} 

\end{enumerate}

The maps (\ref{nearb10}), (\ref{nearb12}) and (\ref{nearb13}) can now 
be treated by KAM-type theorems (\cite{Moser}, p.52 or 
\cite{Douady}, p.III-8). In this way, we obtain the following result: 

\propsep

\noindent 
{\bf Theorem 1}
{\it 
Consider a magnetic billiard in a domain with $C^k$ boundary, $k\geq 6$,
such that $0<\rmin\leq\rho(s)\leq\rmax\leq\infty$.
Let $\eps\in\bbbr_+^*$ and assume that one of the following is true
\begin{enumerate}

\item $0<\mu<\infty$, $\eta=\pi-\th$, $L=\Lb+2\pi\mu$, $\sigma=-1$;

\item $\rmax(1+\eps)\leq\mu<\infty$, $\eta=\th$, 
$L=2\pi\mu-\Lb$, $\sigma=+1$;

\item $0<\mu\leq\rmin(1-\eps)$, $\eta=\th$, $L=\Lb-2\pi\mu$, $\sigma=-1$.

\end{enumerate}
Then, there exists $\eps_1>0$ (depending on $\mu$ and $k$) such that, if 
$\omega\in[0,\eps_1)$ and satisfies the diophantine conditions
\begin{equation}
\abs{\frac{\omega}{L}-\frac{p}{q}}\geq\gamma q^{-\nu}
\label{th1_1}
\end{equation}
for some $\gamma, \nu\in\bbbr_+^*$ and for any $\frac{p}{q}\in\bbbq$, 
there is an invariant curve of the form
\begin{eqnarray}
s & = & \xi + V(\xi)
\nonumber \\
\eta & = & \frac{\omega}{2\mu} + U(\xi),
\label{th1_2}
\end{eqnarray}
where $U,V\in C^1$, $V(\xi+L)=V(\xi)+\Lb-L$, $U(\xi+L)=U(\xi)$.
The induced map on this curve has the form
\begin{equation}
\xi\mapsto\xi+\sigma\omega.
\label{th1_3}
\end{equation}
}

This theorem confirms our observation that invariant curves exist in 
three cases:
\begin{enumerate}
\item Near $u=+1,\;(\th=\pi)$, for all values of the magnetic field 
(see the upper parts of fig.\ref{fig_mapell}d-f). They correspond to 
backward skipping trajectories, the curvature of which is opposite to 
the curvature of the boundary.

\item Near $u=-1,\;(\th=0)$, in weak magnetic field ($\mu\geq\rmax(1+\eps)$,
see the lower part of fig.\ref{fig_mapell}d). They correspond to 
forward skipping trajectories (fig.\ref{fig_ellipse}d), which are curved 
away from the boundary.

\item Near $u=-1,\;(\th=0)$, in strong magnetic field ($\mu\leq\rmin(1-\eps)$,
see the lower part of fig.\ref{fig_mapell}f). They correspond to 
backward skipping trajectories (fig.\ref{fig_ellipse}f).

\end{enumerate}

For intermediate values of the magnetic field, there seem to be no 
tori near $u=-1$. Although we have a qualitative understanding of the 
origin of this phenomenon, we are not able to prove the existence of 
a stochastic component of positive measure in this region. This remains
an open problem.

On the other hand, our theorem implies that magnetic billiards in 
smooth (i.e. $C^6$) convex domains can never be ergodic. Unlike in zero 
field, this remains true even when the curvature of the boundary is 
allowed to vanish. In fact, equation (\ref{nearb9}) suggests that 
invariant curves exist near $u=+1$ even when the boundary has slightly 
concave parts, i.e. $\kappa(s)\geq -\kappa_0$, $\kappa_0>0$, and for 
$\mu\leq\frac{1}{\kappa_0}-\eps$. In that case, however, the 
tori near $u=-1$ do not necessarily survive. We will come back 
to that point in section \ref{sec_neg}.


\subsection{The strong field limit}
\label{sec_strongfield}


The limit $\mu\rightarrow 0$ is in some sense singular: indeed, for small
$\mu$, the distance between successive collisions with the boundary is 
of order $\mu$, so that one could get the impression that the particle 
follows the boundary more and more slowly when $\mu$ decreases, and sticks
to the wall when $\mu=0$, violating the conservation of energy.
Of course, this is only an artificial effect, due to the fact that 
we use the bouncing map instead of the flow to describe the dynamics.
Indeed, the ``time of flight'' between successive collisions is also of 
order $\mu$ (more precisely, it is equal to $\mu\psi$, where $\psi$ 
is given in fig.\ref{fig_arc}), so that during a given time interval, 
the particle hits the boundary $\Order{1/\mu}$ times, traveling a finite 
distance.

Thus, we have to be careful when we apply perturbation theory.
It is possible to use (\ref{nearb4}) to analyze the limit 
$\mu\rightarrow 0$. However, we give in appendix \ref{app5} an 
alternative approach that we find more instructive. The idea is to 
show that although the map is not twist, it can be described by generating 
functions, in fact two of them, with suitable matching at $\chi=\pi/2$.
These functions are of the form 
$G^\pm(s_0,s_1,\mu) = \pm\mu g_0(\sigma) + \mu^2 g_1^{\pm}(\sigma,s_0,\mu)$, 
where $\sigma = \frac{s_0-s_1}{\mu}$. 
Replacing this in (\ref{gen1}), we see that $\sigma$ and $u$ are 
``slow'' variables which evolve on a much longer time scale than $s$. 
This expresses the geometrical idea that for a short skip, the boundary 
is close to an arc of circle, for which $\sigma$ and $u$ are constants.
It is however necessary to use the variable $\th$ instead of $u$ if we 
wish that the map be smooth at the boundaries of the phase cylinder.
Finally, we obtain the following result:

\propsep

\noindent 
{\bf Proposition 3}
{\it 
Consider a magnetic billiard with a convex, $C^k$ boundary, $k \geq 3$.
For small enough $\mu$, the bouncing map is $C^{k-1}$ and has the form:
\begin{eqnarray}
s_1 &=& s_0 - 2\mu\sin\th_0 + \mu^2\sin\th_0 \, a(s_0,\th_0,\mu)
\pmod{\Lb},
\nonumber \\
\th_1 &=& \th_0 + \mu^2 \sin^2\th_0 \, b(s_0,\th_0,\mu).
\label{prop3_1}
\end{eqnarray}
The functions $a\in C^{k-2}$ and $b\in C^{k-3}$ are uniformly bounded for 
$s\in\bbbr,0\leq\th\leq\pi$, $\Lb$-periodic in $s$, and admit the
expansions
\begin{eqnarray}
a(s,\th,\mu) &=& \sum_{i=0}^{k-3} a_i(s,\th)\mu^i + \Order{\mu^{k-2}}, 
\nonumber \\
b(s,\th,\mu) &=& \sum_{j=0}^{k-4} b_j(s,\th)\mu^j + \Order{\mu^{k-3}}
\;\;\;(k\geq 4),
\label{prop3_2}
\end{eqnarray}
where the first terms are
\begin{eqnarray}
a_0 &=& -2\cos\th\kappa(s), \nonumber \\
a_1 &=& -\frac{2}{3}\brak{(1+2\cos 2\th)\kappa(s)^2-\sin2\th\,\kappa'(s)},
\nonumber \\
b_0 &=& \frac{2}{3}\kappa'(s), \nonumber \\
b_1 &=& \frac{4}{3}\cos\th\kappa(s)\kappa'(s) - \frac{2}{3}\sin\th\kappa''(s).
\label{prop3_3}
\end{eqnarray}
}

\propsep

Equation (\ref{prop3_1}) shows that the bouncing map in strong magnetic field
behaves like a perturbed integrable map. To zeroth order in $\mu$, it 
reduces to identity. To first order in $\mu$, it still has the integrable form
$\th_1=\th_0, \, s_1 = s_0 + \mu \Omega(\th_0)$, where $\Omega(\th)=-2\sin\th$.
However, in contrast with usual integrable systems, the frequency $\Omega$ 
is multiplied with the small parameter $\mu$.
As for the factors $\sin\th$ occurring in (\ref{prop3_1}), they assure 
that the boundaries $\th = 0,\, \pi$ are fixed.

The behaviour when $\mu\rightarrow 0$ can be understood in the following way:
fix some positive $\mu_0$ and define $s=\mu_0 \ph$, $\tilde{a}(\ph,\th,\mu) = 
a(\mu_0 \ph,\th,\mu)$, $\tilde{b}(\ph,\th,\mu) = b(\mu_0 \ph,\th,\mu)$. 
Consider the 
2-parameter family of maps $T_{\mu_0,\mu}$:
\begin{eqnarray}
\ph_1 & = & \ph_0 - 2\sin\th_0 + \mu\sin\th_0\,\tilde{a}(\ph_0,\th_0,\mu)
\pmod{\frac{\Lb}{\mu_0}}, 
\nonumber \\
\th_1 & = & \th_0 + \mu^2\sin^2\th_0\,\tilde{b}(\ph_0,\th_0,\mu).
\label{str1}
\end{eqnarray}

The map $T_{\mu_0,0}$ is integrable, $\th$ being a constant of motion, 
and it takes about $n(\th)=\frac{\Lb}{2\mu_0\sin\th}$ iterations for $\ph$
to turn once around phase space. $T_{\mu_0,\mu_0}$ is equivalent to 
(\ref{prop3_1}), and one can expect that when $\mu_0$ is small, some features
of the integrable map remain, in particular the particle completes one turn 
after roughly $n(\th)$ bounces. Using (\ref{A2}) and (\ref{prop3_1}), 
we obtain that $\psi=2\chi=2\pi-2\th_0+\Order{\mu}$, so that the time 
necessary for one revolution is of order 
$t\sim n(\th)\mu_0\psi\sim\Lb\frac{\pi-\th}{\sin\th}$, 
which does not depend on $\mu$ to lowest order.
Note an important difference between the skipping regimes $u\sim +1$ 
and $u\sim -1$: in the first case, $t\sim\Lb$, whereas in the second 
case $t\sim\Lb\frac{\pi}{\th}$ diverges when $u\rightarrow -1$. 
The variation of $\th$ during $t$ is of order
$n(\th)(\th_1-\th_0)\sim\mu_0\sin\th$, which shows that our approximation 
is consistent, since $\th$ is an adiabatic invariant on the time scale $t$.

We have the choice between two different perturbative techniques to make 
these observations mathematically precise.
If we exclude some neighborhood of $\th=\frac{\pi}{2}$, where the 
frequency $\Omega(\th)=-2\sin\th$ has a vanishing derivative, we can apply 
Moser's theorem (\cite{Moser}, p.52): if the boundary is $C^k, k\geq 6$, 
then there exist invariant curves for every frequency satisfying some 
diophantine condition\footnote{One advantage of Moser's theorem is that the 
map need not preserve an annulus.}. In this way however, we cannot describe 
trajectories starting with a very small tangential velocity 
($\th\sim\frac{\pi}{2}$).
In order to do this, we will apply another technique to study 
(\ref{prop3_1}), coming from adiabatic theory. We will prove in 
section \ref{sec_theo} the following theorem on a class of maps 
of the annulus:

\propsep

\noindent 
{\bf Theorem 2 (An adiabatic theorem for maps)}

\noindent
{\it 
Let the map $T:\pim0\mapsto(\ph_1,I_1)$,
\begin{eqnarray}
\ph_1 & = & \ph_0 + \mu c(I_0)\brak{\Omega\im0+\mu\alpha\pim0}
\nonumber \\
I_1 & = & I_0 + \mu^2 c(I_0)^2 \beta\pim0,
\label{th1}
\end{eqnarray}
be defined on the set 
$E = \set{(\ph,I,\mu) \mid \ph\in\bbbr,0\leq I\leq 1,0\leq\mu\leq\mu^*}$.
Assume that $\alpha$ and $\beta$ are periodic functions of $\ph$ with 
period $1$, $c(0)=c(1)=0$, and $\Omega(I,0)=\W0\neq 0$.

Then, there exists a change of variables
$(\ph,I)\mapsto(\psi=\ph+\mu f(\ph,I,\mu),J=I+\mu c(I)g(\ph,I,\mu))$,
preserving the square $[0,1]\times [0,1]$, 
where $f$ and $g$ are periodic with period $1$ in $\ph$, such that
\begin{enumerate}

\item If $T$ is analytic in a complex neighborhood of $E$, then
\begin{eqnarray}
\psi_1 & = & \psi_0 + \mu c(J_0)
\brak{\bar{\Omega}\jm0+\e^{-1/C\mu}\bar{\alpha}\pjm0}
\nonumber \\
J_1 & = & J_0 + \mu^2 c(J_0)^2
\brak{\Theta\jm0+\e^{-1/C\mu}\bar{\beta}\pjm0}.
\label{th2}
\end{eqnarray}

\item If moreover $T$ preserves the measure $c(I)\rho (\ph,I,\mu)d\ph dI$, 
where $\rho(\ph,I,0)=1$, then
\begin{equation}
J_1 = J_0 + \e^{-1/C\mu}c(J_0)^2\bar{\beta}\pjm0.
\label{th3}
\end{equation}

\item If $\alpha$, $\beta$ and $\Omega$ are $C^k$ 
and $c$ is $C^{k+1}$ in $E$, then
\begin{eqnarray}
\psi_1 & = & \psi_0 + \mu c(J_0)
\brak{\bar{\Omega}\jm0+\mu^{k+1}\bar{\alpha}\pjm0}
\nonumber \\
J_1 & = & J_0 + \mu^2 c(J_0)^2
\brak{\Theta\jm0+\mu^k\bar{\beta}\pjm0},
\label{th4}
\end{eqnarray}
respectively
\begin{equation}
J_1 = J_0 + \mu^{k+2} c(J_0)^2\bar{\beta}\pjm0
\label{th5}
\end{equation}
if $T$ preserves the measure $c(I)\rho(\ph,I,\mu)d\ph dI$.

\end{enumerate}
}

\propsep

Equation (\ref{th1}) describes a map from the annulus $0\leq I\leq 1$ 
into itself (the choice of 0, 1, is arbitrary and can be changed by scaling). 
Indeed, the factor $c(I)$, which vanishes at $I=0,\,1$, 
assures that the boundaries of the annulus are fixed, and for small 
enough $\mu$, the interior of the annulus is invariant.
In (\ref{th1}), the terms containing the phase $\ph$, which make the 
map non-integrable, are of order 
$\mu^2$. The theorem states that a suitable change of variables 
decreases this order to $\mu^{k+2}$ if the map is $C^k$, and to the 
exponentially small order $\e^{-1/C\mu}$ if the map is analytic.

The method of the proof is similar to that used by Nehoroshev and Neishtadt 
(see \cite{Arnold2}, p.163 and references therein) in the case of 
differential equations. The basic idea is to make successive changes 
of variables that each decrease by one unit the order of the terms 
containing the phase $\ph$. 
If the map is $C^k$, one can repeat this procedure $k$ times. If it is 
analytic, divergence prevents us from applying the procedure infinitely 
often, but it is possible to do it a large number of times, of 
order $\frac{1}{\mu}$, so that the terms containing the phase 
become exponentially small.

We can now apply the theorem to the bouncing map (\ref{prop3_1}) 
of billiards in strong magnetic field, where $c(\th)=\sin\th$.
Because of proposition 1, the map preserves the measure 
$\sin\th ds d\th$, and can thus be transformed into the form
(\ref{th3}) or (\ref{th5}). If we write $J_i = J(s_i, \th_i)$, 
we obtain

\propsep 

\noindent
{\bf Corollary}
{\it
Consider a magnetic billiard in a convex domain. If the boundary is 
$C^k,k\geq 3$, then there exists a function $J^{(k)}(s,\th)$ such that 
$J^{(k)}_1=J^{(k)}_0+\Order{\mu^{k-1}}$.
If the boundary is analytic (in the sense that $\kappa(s)$ can be analytically 
continued to a complex neighborhood of the real axis), then there exists a 
function $J(s,\th)$ such that $J_1=J_0+\Order{\e^{-1/C\mu}}$.
In any case, this function can be written in the form 
$J=\th+\mu\sin\th g(s,\th,\mu)$. 
}

\propsep

If the boundary is $C^k,k\geq 3$, this result implies that after $n$ bounces,
we have $J_n = J_0 + n\Order{\mu^{k-1}}$, and hence if $n = \Order{\mu^{2-k}}$, 
$J_n = J_0 + \Order{\mu}$. In other words, the quantity $J$ is a 
``quasi-invariant'' or ``adiabatic invariant'' which varies of an amount of 
order $\mu$ for $\Order{\mu^{2-k}}$ bounces, i.e. during a time of order 
$\mu^{3-k}$. In the analytic case, $J$ varies on a time scale growing 
exponentially with the magnetic field.
In contrast with KAM theorems, which show the existence of {\em exact} 
invariants for {\em some} initial conditions, the adiabatic theorem 
shows the existence of {\em approximate} invariants for {\em all}
initial conditions.

\begin{figure}
\centerline{\psfig{figure=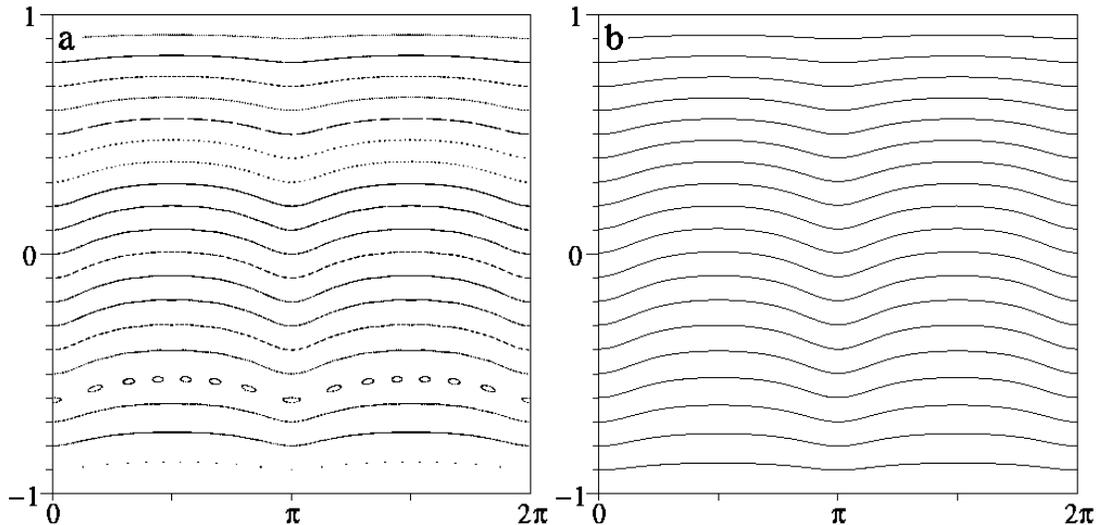,height=70mm,clip=t}}
\caption{Adiabatic invariant in the strong field limit:
	(a) some orbits of the billiard in an ellipse for 
	    $\mu=0.3$, $\lambda=1.5$ (i.e. $\rmin=\frac{2}{3}$),
	(b) level lines of the adiabatic invariant (\protect\ref{str2}), 
	    which are given by
	    $\th = J - \mu \sin J\brak{\frac{1}{3}\kappa(s) + \frac{1}{9}\mu 
	    \cos J\kappa(s)^2} + \Order{\mu^3}$.}
\label{fig_adiab}
\end{figure}

In fact, the theorem not only shows the existence of quasi-invariants, it 
also enables us to compute them up to the first orders. For example, after 
2 changes of variables (given by (\ref{pth1}), (\ref{pth2}) and 
(\ref{pth3})), we obtain
\begin{equation}
J^{(5)}(s,\th,\mu) = \th + 
\mu\sin\th\brak{\frac{1}{3}\kappa(s) + \frac{2}{9}\mu\cos\th\kappa(s)^2}.
\label{str2}
\end{equation}
To zeroth order in $\mu$, we recover that $J^{(3)} = \th$ is an 
adiabatic invariant on a time scale of order 1.
To first order in $\mu$, $J^{(4)}$ corresponds to the invariant (\ref{convex1}) 
of \cite{BeRo}\footnote{Namely, $K = f(J^{(4)}) + \Order{\mu^2}$,
where $f(J) = -\cos J(1+2\sin^2 J)\sin^{-3}J$.}, which is valid for a time 
of order $\mu^{-1}$. 
As shown in figure \ref{fig_adiab}, the orbits are very 
close to the level lines of $J^{(5)}$, even when $\mu$ is of the 
same order as $\rmin$. The dynamics on the quasi-invariant curve 
$J(s,\th) = J_0$ is approximated by $\psi_1 = \psi_0 + 
\mu c(J_0) \bar{\Omega}(J_0,\mu)$.

In section \ref{sec_neg}, we will say a word on adiabatic invariants 
for billiards in non-convex boundaries.
Quasi-invariants can also be used to study billiards with piecewise smooth 
boundaries. As long as the particle hits the same smooth piece of the boundary,
the corresponding quasi-invariant changes slowly, so that one can estimate 
the outcoming velocity. 
However, as soon as the particle passes from one smooth piece to another one,
a jump of the value of $J$ occurs. This is particularly clear in the 
stadium, where $J=\th$ is even an exact invariant on each straight line 
and arc of circle. The jumps of $\th$ occurring at the junction points
allow the particle to explore a large fraction of phase space (see also
\cite{Ro}). Thus, billiards in insufficiently smooth, convex domains do 
{\em not} become integrable in the strong field limit 
(as suggested in \cite{MBG}).


\section{More chaotic billiards}
\label{sec_nonsmooth}


\begin{figure}
 \centerline{\psfig{figure=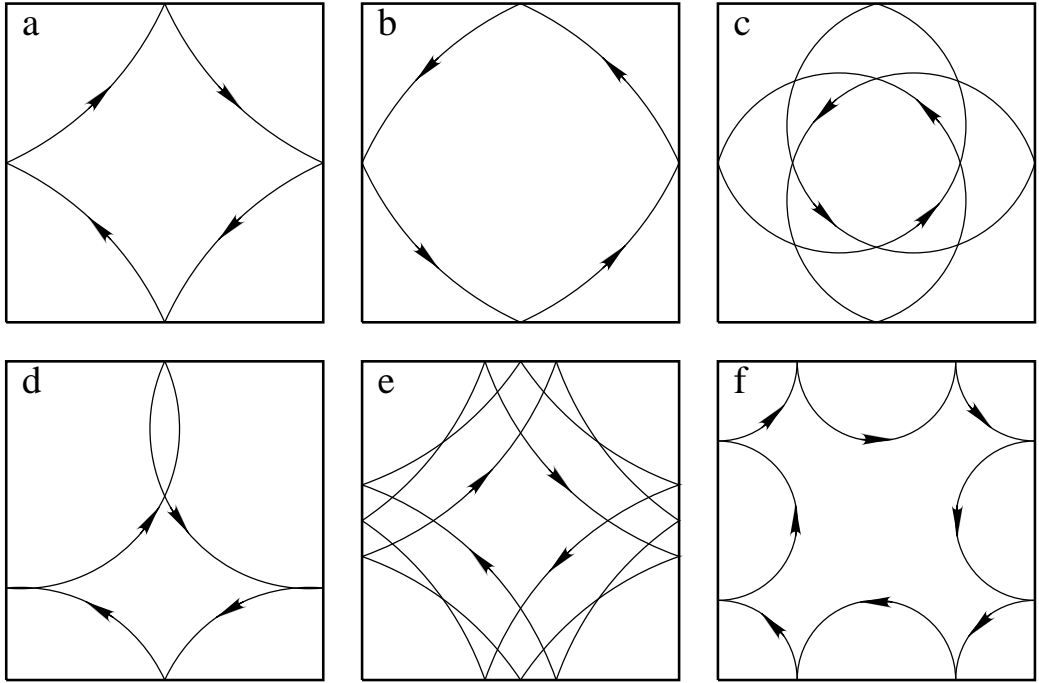,height=90mm}}
 \caption{Periodic trajectories in a square.Trajectories of period 4:
	 (a)~type I, (b)~type II, (c)~type III,
	 (d)~asymmetric trajectory at $\mu = 0.502$, bifurcating with type I
	 at $\mu = \half$.
	 (e)~Trajectory of period 12 which bifurcates with type I when 
	 $\mu=\protect\frac{\protect\sqrt5}{2}$.
	 (f)~Stable symmetric trajectory of period $4k$, $k = 2$ for 
	 $\mu = \protect\frac{1}{4}$.}
\label{fig_square}
\end{figure}

 
\subsection{The billiard in a square}
\label{sec_square}


Billiards in convex polygonal domains, in the presence of a magnetic 
field, can be expected to show a chaotic tendency, because we will always 
be in the regime $0=\rmin<\mu<\rmax=\infty$. 

We have concentrated our study to the case of a square of side length $1$. 
This billiard is integrable in zero field, where every periodic orbit is
parabolic. In fact, the periodic orbits occur in families, which can be 
indexed by the slope of the trajectory, and whose members can be indexed 
by the arclength of any collision with the boundary.
 
\begin{figure}[t]
 \centerline{\psfig{figure=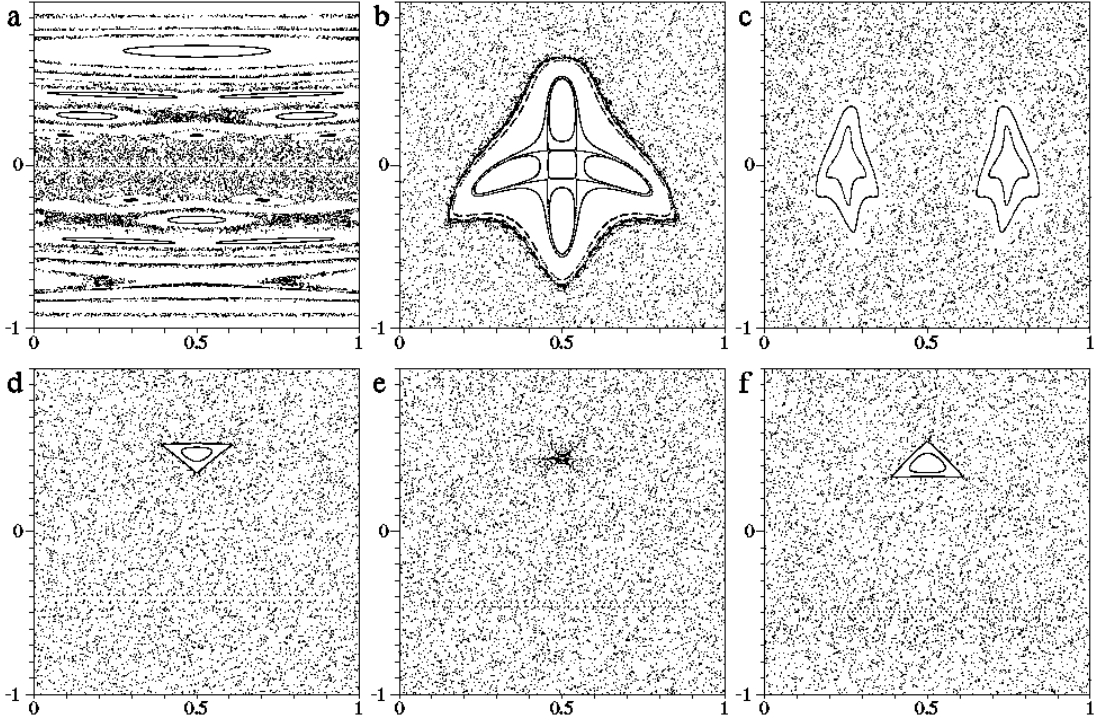,height=95mm,clip=t}}
 \caption{Phase portraits of the billiard in a square.
	 (a)~$\mu = 30$, (b)~$\mu = 0.502$, (c)~$\mu = \protect\frac{1}{4}$.
	 ``Squeeze effect'' bifurcation around $\mu = 
	 \protect\frac{\protect\sqrt{5}}{2} = 1.118\ldots$:
	 (d)~$\mu = 1.25$, (e)~$\mu = 1.118$, (f)~$\mu = 1$.}
\label{fig_phasquare}
\end{figure}

In low magnetic field, only isolated orbits subsist, some of which are 
hyperbolic, and some elliptic (fig.\ref{fig_phasquare}a)\footnote{The
first correction to the zero-field generating function goes like the 
magnetic flux through the trajectory, i.e. the signed area enclosed by 
the trajectory.
For orbits whose period is a multiple of $4$, it can be seen that this area 
varies quadratically with the arclength indexing the orbit.}.
 When the field increases, most of the 
latter vanish or get unstable, and structure of phase space is dominated 
by symmetric period-4 orbits (fig.\ref{fig_square}a-c): type I exists for 
$\mu \geq \frac{1}{\sqrt{8}}$, type II for $\mu > \half$, and type III for 
$\frac{1}{\sqrt{8}} \leq \mu < \half$.

Using (\ref{map2}), we find that type II and III orbits are hyperbolic, 
except for $\mu = \frac{1}{\sqrt{8}}$, where type III and type I undergo 
saddle-node bifurcation. For type I orbits, we find that the stability 
matrix satisfies
\begin{equation}
 \half \Tr S_4 = 8t^4 - 8t^2 + 1, \;\;\; 
 t = \half \Tr S_1 = \frac{1-\sqrt{8\mu^2 - 1}}{1+\sqrt{8\mu^2 - 1}},
\label{square1}
\end{equation}
so that these are elliptic, except for $\mu = \frac{1}{\sqrt{8}}$ 
(saddle-node bifurcation), $\mu = \half$ (bifurcation with 4 elliptic and 
4 hyperbolic asymmetric orbits of period 4, fig.\ref{fig_square}d, 
\ref{fig_phasquare}b), and 
$\mu = \frac{\sqrt{3}}{2\sqrt{3\pm 2\sqrt{2}}}$ (no bifurcation).

\begin{figure}[t]
 \centerline{\psfig{figure=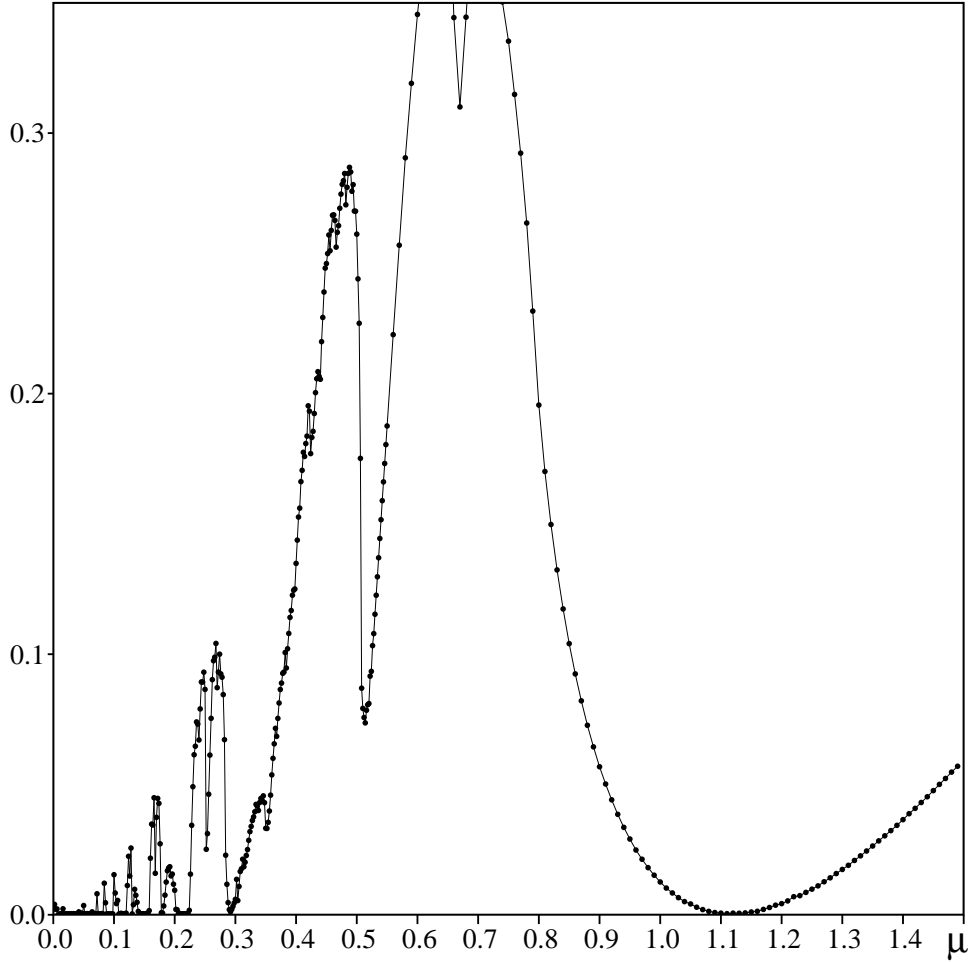,height=130mm}}
 \caption{Numerically estimated proportion of regular component in phase
 	 space for the 
	 billiard in a square, as a function of $\mu$. Actually, we show the 
	 relative area of phase space not occupied by one chaotic orbit,
	 with a precision of $4 \cdot 10^{-6}$.}
\label{fig_stoc}
\end{figure}

An interesting resonance phenomenon occurs when $\mu = \frac{\sqrt{5}}{2}$.
$S_4$ is then a cubic root of $\bbbone$, and type I orbit becomes unstable 
because of a ``squeeze effect'' bifurcation (\cite{Arnold}, p.392) with a 
hyperbolic orbit of period 12 (fig.\ref{fig_square}e, 
\ref{fig_phasquare}d-f)\footnote{The same kind of bifurcation affects the 
small diametral orbit in the ellipse when $\lambda = 2$, $\forall \mu > 1$.}.
Numerical simulations show no other stable periodic orbits in this case, 
leading us to the conjecture that the billiard in a square is ergodic 
for this particular value of $\mu$. Figure \ref{fig_stoc} illustrates this 
phenomenon by showing a numerical estimate of the size of regular components 
of phase space, which vanishes for $\mu = \frac{\sqrt{5}}{2} \simeq 1.118\ldots$.

There are no orbits of period 4 when $\mu < \frac{1}{\sqrt{8}}$. However, 
stable orbits of higher period exist for arbitrarily high values of
the magnetic field. Figure \ref{fig_square}f shows a symmetric trajectory 
of period 8, doing 2 bounces on each side of the square. We can construct 
similar trajectories of period $4k$, $k \geq 1$, reflected $k$ times on 
each side. Detailed calculations (see appendix \ref{app3}) show that such 
orbits are stable for almost all $\mu$ in some interval containing 
$\frac{1}{2k}$, but these intervals do not overlap when $k>2$, leading us to 
the assumption that mixed and ergodic behaviour alternate for decreasing $\mu$.

 
\subsection{Boundaries with negative curvature}
\label{sec_neg}


In Sinai billiards, whose boundaries have negative curvature, it is known 
that all periodic orbits are unstable in zero field (as is suggested by the 
fact that $\Tr(DT) \leq -2$ if we put $\chi = 0$ in (\ref{map2})). This is no 
longer true in non-zero field, because the curved trajectories, although 
locally dispersed, may converge again.

\begin{figure}
 \centerline{\psfig{figure=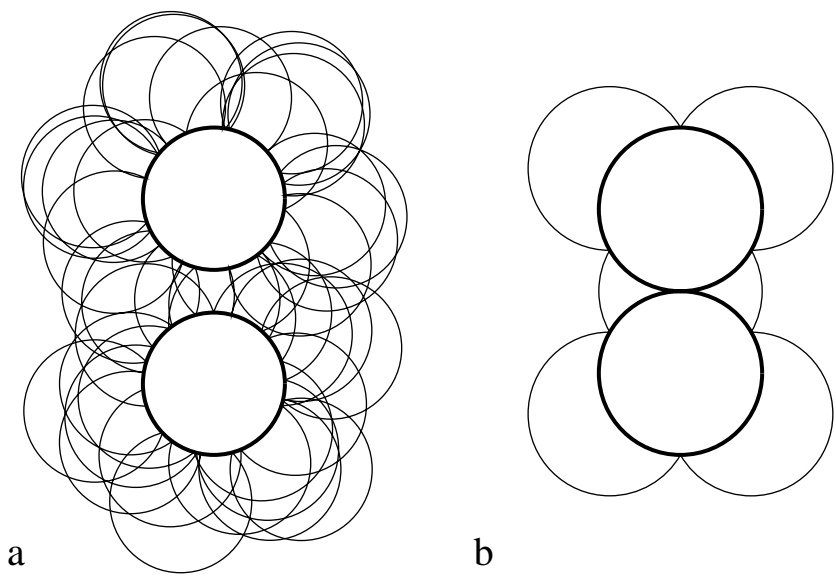,height=80mm}}
 \caption{The billiard outside two circles: 
	 (a) chaotic trajectory, (b) one of the few stable trajectories.}
\label{fig_circles}
\end{figure}

Let us first consider billiards {\em outside} smooth convex boundaries
with extremal radii of curvature $\rmin$ and $\rmax$.
If we fix a trajectory of the outside billiard, and complete every arc 
of trajectory to a full circle, we obtain its ``dual trajectory'' as the 
set of all complementary arcs. If the dual trajectory never crosses the
boundary, then it corresponds to a real trajectory of the inside billiard,
since the law of specular reflection is satisfied by construction.
A sufficient condition for this to be true is that any circle of radius 
$\mu$ can intersect the boundary at most twice. In this section, let us 
call this the ``$\mu$-intersection property''. This property is actually 
granted in two cases: in strong field $\mu\leq\rmin$ (as we prove in 
lemma 3 of appendix \ref{app5}), and in weak field $\mu\geq\rmax$ 
(as can be proven in a similar way). In these cases, to any orbit 
$(\ldots(s_{-1},u_{-1}),(s_0,u_0),(s_1,u_1)\ldots)$ of the outside 
billiard, corresponds the orbit
$(\ldots(-s_1,-u_1),(-s_0,-u_0),(-s_{-1},-u_{-1})\ldots)$
of the inside billiard, and reciprocally.
Thus, the billiards inside and outside the boundary are perfectly 
equivalent, so that we can apply our results of section \ref{sec_convex}
on existence of invariant curves and adiabatic invariants (theorem 1, 
proposition 3 and the corollary of theorem 2). In particular,
the billiard outside a circle is integrable.

For intermediate values of the magnetic field $(\rmin<\mu<\rmax)$, 
the dual of any outside trajectory will not necessarily be an inside 
trajectory. However, the equivalence is true for some special 
trajectories, namely those which are sufficiently close to $u=+1$ for 
the inside billiard (and remain so because of theorem 1).

Consequently, we predict that if we compare the phase portraits of the 
billiards inside and outside a given smooth convex curve, they will be the
same (up to an inversion) for $\mu\leq\rmin$ and $\mu\geq\rmax$.
When $\rmin<\mu<\rmax$, the region $u\sim 1$ of the interior portrait 
will be the same as the region $u\sim -1$ of the exterior portrait, 
but other parts of them will be different.

Let us consider next the case where the boundary $\dQ$ has both 
convex and concave parts, but a bounded curvature: $\abs{\kappa(s)}
\leq\kappa_0 = \frac{1}{\rmin} \forall s$. We can argue by the means
of geometrical properties that such a billiard must possess invariant 
tori and adiabatic invariants in sufficiently strong magnetic field, 
if $\dQ$ is smooth enough.
It should be clear that the interior of the domain has to be connected, 
since otherwise we would have several independent billiards. 
We can thus define $r^*$ such that the $\mu$-intersection property 
is satisfied for $\mu\leq r^*$ \footnote{Note that the bouncing map is 
continuous iff the $\mu$-intersection property is satisfied. Indeed, 
discontinuities by tangency exist iff one can construct a Larmor circle 
tangent to the boundary, and intersecting it at two other points}. 
This radius has to be positive 
but can be arbitrarily small (consider billiards shaped like
sand-glasses or peanuts).

In section \ref{sec_nearbound}, we already mentioned that the expression
(\ref{nearb9}) of the bouncing map near $\th=\pi$ can still be valid 
if $\mu<\rmin(1-\eps)$, namely if $\th$ is sufficiently close to $\pi$ 
that the trajectory remains within a distance $2 r^*$ of the boundary.
Thus, invariant curves still exist near $\th=\pi$. Furthermore, we 
note that the only point in the proof of proposition 3 where we 
use the convexity of the boundary is to show the $\mu$-intersection 
property. This is actually satisfied for $\mu\leq r^*$, so that we 
conclude that proposition 3 and its consequences on adiabatic invariants
remain true in the present case, for sufficiently small $\mu$.

Finally, we also studied the billiard outside 2 circles of radius 1, 
centered at $(0, \pm \lambda)$
(fig.\ref{fig_circles}). If $\lambda > 1$, it is important to note that 
$M_1$ defined in section \ref{sec_def} is not the only integrable component 
of phase space. Indeed, trajectories with 
$u>u^* = \min\left\{1, 2\lambda-1-\frac{2\lambda(\lambda-1)}{\mu}\right\}$
always touch the same circle, so that $u$ is a constant of motion. 
In particular, when $\mu < \lambda-1$, the billiard is integrable.
When $\mu > \lambda-1$, the map is discontinuous whenever trajectories become 
tangent to the boundary, and numerical simulations show that the component 
of phase space $u<u^*$ is almost filled with a stochastic sea with positive 
Liapunov exponents. However, one can find stable periodic orbits for particular
values of the parameters (see appendix \ref{app6}), so that the billiard is not always ergodic.


\section{Proof of the adiabatic theorem for maps}
\label{sec_theo}


 
\subsection{Construction of the change of variables}

 
The change of variables is constructed as the composition of several
elementary substitutions, that each increase by one the order of the terms 
containing the phase. The number of these substitutions is $k$ if 
the map is $C^k$ and of order $\brak{1/\mu}$ if the map is analytic.

We assume by induction that after $N$ steps, the map reads
\begin{eqnarray}
\ph_1 & = & \ph_0 + \mu c(I_0)
\brak{\Omega_N\im0+\mu^{N+1}\alpha_N\pim0}
\nonumber \\
I_1 & = & I_0 + \mu c(I_0)^2
\brak{\mu\Theta_N\im0+\mu^{N+1}\beta_N\pim0},
\label{pth1}
\end{eqnarray}
where we write $(\ph,I)=(\ph^{(N)},I^{(N)})$ for brevity,
and $\Omega_N(I,\mu)=\W0+\mu\bar{\Omega}_N(I,\mu)$. We can 
assume that $\avrg{\alpha_N}=\avrg{\beta_N}=0$, 
where $\avrg{F}=\int_0^1 Fd\ph$ denotes the average of $F$.

The change of variables $(\ph,I)\mapsto(\psi,J)=(\ph^{(N+1)},I^{(N+1)})$
is defined by
\begin{eqnarray}
\psi & = & \ph + \mu^{N+1} f(\ph,I,\mu)
\nonumber \\
J & = & I + \mu^{N+1} c(I) g(\ph,I,\mu),
\label{pth2}
\end{eqnarray}
where
\begin{eqnarray}
g(\ph,I,\mu) & = & \int_0^\ph -\frac{1}{\W0}\beta_N(\ph',I,\mu)d\ph',
\nonumber \\
g_0(\ph,I,\mu) & = & g(\ph,I,\mu) - \avrg{g}(I,\mu),
\nonumber \\
f(\ph,I,\mu) & = & \int_0^\ph 
\brak{c'(I)g_0(\ph',I,\mu)-\frac{1}{\W0}\alpha_N(\ph',I,\mu)}d\ph'.
\label{pth3}
\end{eqnarray}

We will use the following properties:
\begin{enumerate}

\item If $F(\ph)\in C^1$, then
\begin{eqnarray}
\int_{\ph_0}^{\ph_1}F(\ph)d\ph 
& = & \mu c(I_0)\brak{\Omega_N(I_0,\mu)F(\ph_0)+\mu r_1\pim0}
\nonumber \\
& = & \mu c(I_0)\brak{\W0 F(\ph_0)+\mu r_2\pim0}.
\label{pth4}
\end{eqnarray}

Indeed, carrying out the change of variables $\ph=\ph_0+\mu c(I_0)x$,
one obtains
\[
\int_{\ph_0}^{\ph_1}F(\ph)d\ph 
= \mu c(I_0)\brak{\int_0^{\Omega_N}F(\ph_0)dx+
\mu c(I_0)\int_0^{\Omega_N}F'xdx+
\int_{\Omega_N}^{\Omega_N+\mu^{N+1}\alpha_N}Fdx}.
\]

\item If $F(I)\in C^1$, then 
\begin{equation}
F(I_0) = F(J_0) + \mu^{N+1}c(I_0)r_3\pim0,
\label{pth6}
\end{equation}
where $r_3=-g\pim0 F'(I)$. In particular, $c(I_0)=c(J_0)R_0\pim0$, 
where $R_0=\brak{1-\mu^{N+1}gc'}^{-1}$.

\end{enumerate}

Substituting (\ref{pth1}) in (\ref{pth2}), we get
\begin{eqnarray}
J_1-J_0 & = & \mu^{N+1}\brak{c(I_1)g\Pim1 - c(I_0)g\pim0} +
\nonumber \\
 &  & + \mu^2 c(I_0)^2 \brak{\Theta_N\im0 + \mu^N\beta_N\pim0}.
\label{pth7}
\end{eqnarray}

Now, using the above properties, we obtain
\begin{eqnarray}
c(I_1)g\Pim1 - c(I_0)g(\ph_1,I_0,\mu) & = & 
(I_1-I_0)\pd{I}\brak{c(I)g(\ph_1,I,\mu)}
\nonumber \\
 & = & \mu^2 c(I_0)^2 R_1\pim0,
\nonumber \\
c(I_0)g(\ph_1,I_0,\mu) - c(I_0)g\pim0 & = &
-\frac{1}{\W0}c(I_0)\int_{\ph_0}^{\ph_1}\beta_N(\ph,I_0,\mu)d\ph 
\nonumber \\
 & = & -\mu c(I_0)^2 \beta_N\pim0 +
\nonumber \\
 &   &  + \mu^2 c(I_0)^2 R_2\pim0,
\nonumber \\
c(I_0)\Theta_N\im0 - c(J_0)\Theta_N\jm0 & = & 
\mu^{N+1}c(I_0)R_3\pim0,
\nonumber \\
c(I_0) - c(J_0) & = & 
\mu^{N+1}c(I_0)r_4\pim0.
\label{pth8}
\end{eqnarray}

Replacing this in (\ref{pth7}), we finally get
\begin{equation}
J_1=J_0 + \mu^2 c(J_0)^2\Theta_N\jm0 
+ \mu^{N+3}c(J_0)^2 \tilde{\beta}_{N+1}\pim0,
\label{pth9}
\end{equation}
where $\tilde{\beta}_{N+1}=R_0^2(R_1+R_2+R_3)+R_0R_4$
and $R_4=\Theta_Nr_4$. The final step is to write 
$\tilde{\beta}_{N+1}\pim0=\hat{\beta}_{N+1}\pjm0$, which can be done 
for small enough $\mu$, as we shall check later, and
$\Theta_{N+1}=\Theta_N+\mu^{N+1}\avrg{\hat{\beta}_{N+1}},
\beta_{N+1}=\hat{\beta}_{N+1}-\avrg{\hat{\beta}_{N+1}}$.

Proceeding in a similar way for $\psi$, we obtain
\begin{eqnarray*}
\psi_1-\psi_0 &=& \mu c(I_0)\Omega_N\im0
\brak{1+\mu^{N+1}c'(I_0)g_0\pim0} + \mu^{N+3}c(I_0)R_5
 \\
c(I_0)\Omega_N\im0 &=& c(J_0)\Omega_N\jm0
+\mu^{N+1}c(I_0)\brak{-\Omega_N\im0 c'(I_0)g\pim0 + \mu R_6}
\end{eqnarray*}
\[
c(I_0)c'(I_0)\Omega_N\im0 \avrg{g}\im0 = 
c(J_0)c'(J_0)\Omega_N\jm0 \avrg{g}\jm0 - \mu c(I_0)R_7,
\]
so that
\begin{equation}
\psi_1=\psi_0+ \mu c(J_0)\hat{\Omega}_N\jm0 
+ \mu^{N+3}c(J_0) \tilde{\alpha}_{N+1}\pim0.
\label{pth11}
\end{equation}
where $\tilde{\alpha}_{N+1}=R_0(R_5+R_6+R_7)$
and $\hat{\Omega}_N=\Omega_N(1-\mu^{N+1}c'\avrg{g})$.
Finally, we proceed in the same way as for $\beta$ in order to 
define $\Omega_{N+1}, \alpha_{N+1}$.

 
\subsection{Bounds and domains of analyticity}

 
In the case where the map is analytic, we define
\begin{equation}
\Gamma(D)=\set{(\ph,I)\mid \abs{\Im\ph},\abs{\Im I}< D,
-D<\Re I<1+D}.
\label{pth13}
\end{equation}
We assume that for $(\ph,I)\in \Gamma(D_N)$ 
and $\abs{\mu}<\mu_0$, (\ref{pth1}) is analytic and satisfies 
the bounds $\abs{\alpha_N},\abs{\beta_N}\leq M_N$,
$\abs{\bar{\Omega}_N},\abs{\Theta_N}\leq W_N$.
We introduce numbers $D_N''<D_N'<D_N$, such that
$(\ph_0,I_0)\in \Gamma(D_N'')$ implies 
$(\ph_1,I_1),(\psi_0,J_0)\in \Gamma(D_N')$.
Thus, $D_N''$ has to satisfy the condition
\begin{equation}
D_N'' \leq D_N' - 
k_0 \max\set{\mu_0+\mu_0^2W_N+\mu_0^{N+2}M_N,\mu_0^{N+1}M_N}.
\label{pth14}
\end{equation}

From now on, the numbers $k_i, \lambda_i, c_i$ will designate constants
which are uniform in $N$ and $\mu$. Using Cauchy's inequality to bound 
the derivatives appearing in the expressions of the $R_i$, we obtain 
that for $(\ph_0,I_0)\in \Gamma(D_N'')$, $\abs{R_i}\leq K_i$, where
\begin{eqnarray}
K_0 & = & \frac{1}{1-\mu_0^{N+1}\lambda_0M_N}
\nonumber \\
K_1 & = & \lambda_1\brak{W_N+\mu_0^NM_N}\frac{M_N}{D_N-D_N'}
\nonumber \\
K_2 & = & \lambda_2\brak{W_N+\mu_0^NM_N+\frac{1}{D_N-D_N'}}M_N
\nonumber \\
K_3 & = & \lambda_3\frac{M_NW_N}{D_N-D_N'}
\nonumber \\
K_4 & = & \lambda_4 M_NW_N
\nonumber \\
K_5 & = & \lambda_5\brak{\frac{W_N+\mu_0^NM_N}{D_N-D_N'} 
 + \mu_0^NM_N + \frac{1+\mu_0W_N}{D_N-D_N'}}M_N
\nonumber \\
K_6 & = & \lambda_6\brak{1+\mu_0^NM_N}\frac{M_NW_N}{D_N-D_N'}
\nonumber \\
K_7 & = & \lambda_7 \mu_0^NM_N\frac{M_NW_N}{D_N-D_N'},
\label{pth15}
\end{eqnarray}
and the new bounds have to satisfy
\begin{eqnarray}
M_{N+1} & \geq & c_1 \max\set{K_0^2\sum_{i=1}^3K_i+K_0K_4, K_0\sum_{i=5}^7K_i}
\nonumber \\
W_{N+1} & \geq & W_N + c_2\mu_0^{N+1}(W_NM_N+M_{N+1}).
\label{pth16}
\end{eqnarray}

Finally, we have to consider the effect of the change of variables
on the domain of analyticity.

{\bf Lemma 1}
{\it
If $F(\ph,I,\mu)$ is analytic for $(\ph,I)\in\Gamma(D)$, then 
$\bar{F}(\psi,J,\mu)=F(\ph,I,\mu)$ is analytic for 
$(\psi,J)\in\Gamma(D-4c_0\mu_0^{N+1}M_N)$.
}

{\bf Proof:}
Write $x=(\ph,I), y=(\psi,J)$. From (\ref{pth2}), we have $y=x+G(x)$,
with $\abs{G_i(x)}\leq c_0\mu_0^{N+1}M_N$. If $\Phi(x,y) = x+G(x)-y$ 
and $\abs{\pd{i}G_j}\leq\eps$, then $\det(\pd{x}\Phi)\geq 1-2\eps-2\eps^2$.
If $y\in\Gamma(D-4c_0\mu_0^{N+1}M_N)$, then $x\in\Gamma(D-3c_0\mu_0^{N+1}M_N)$,
and Cauchy's inequality implies $\eps<1/3$, so that  $\det(\pd{x}\Phi)>0$.
The implicit function theorem implies that we can write $x=\xi(y)$ where 
$\xi$ is analytic, and so is $\bar{F}(y,\mu)=F(\xi(y),\mu)$.\qed

Hence, we must take
\begin{equation}
D_{N+1} \leq D_N''-4c_0\mu_0^{N+1}M_N.
\label{pth17}
\end{equation}

 
\subsection{Evolution of the bounds with $N$}

 
We now choose domains of the form $D_N=D_0-Nd\mu_0$, $D_N'=D_N-d'\mu_0$,
$d'<d$. We wish to show by induction that for a suitable choice of 
$d$ and $d'$ and $\mu_0$ small enough,
\begin{eqnarray}
M_N & = & \frac{M_0}{(2\mu_0)^N},
\label{pth18} \\
W_N & = & W_0 + c_3\parth{1-\frac{1}{2^N}}.
\label{pth19}
\end{eqnarray}

Replacing this in (\ref{pth15}), we get 
$K_0 \leq\brak{1-\mu_0\lambda_0M_0}^{-1}\leq 2$ if 
$\mu_0\leq\brak{2\lambda_0M_0}^{-1}$, and $K_i\leq k_i M_N/d'\mu_0, 
i=1\ldots 7$. Thus, we can take
\begin{equation}
M_{N+1}=c_1 m \frac{M_N}{d'\mu_0} = \frac{M_N}{2\mu_0}
\label{pth20}
\end{equation}
where $d'=2c_1m$ and $m=\max\set{4\sum_{i=1}^3k_i+2k_4, 2\sum_{i=5}^7k_i}$.

To prove (\ref{pth19}), we take $c_3=W_0+c_2M_0$, $\overline{W}=W_0+c_3$ 
and $\mu_0\leq W_0\brak{2c_2M_0\overline{W}}^{-1}$, so that
\[
c_2\mu_0^{N+1}(W_NM_N+M_{N+1})  \leq 
\frac{c_2M_0}{2^N}\brak{\mu_0\overline{W}+\half} 
  \leq  \frac{1}{2^N}\brak{\frac{W_0}{2}+\frac{c_2M_0}{2}}
 = \frac{c_3}{2^{N+1}}.
\]

Finally, it is easy to check that the conditions (\ref{pth14}) and 
(\ref{pth17}) are satisfied if we take 
$d=d'+4c_0M_0+k_0\max\set{1+\mu^*(\overline{W}+M_0),M_0}$.

The last step in order to prove (\ref{th2}) is to take
\begin{equation}
N = N(\mu_0) = \brak{\frac{D_0}{d\mu_0}}.
\label{pth21}
\end{equation}

In this way, $W_{N(\mu_0)}$ is bounded, $D_{N(\mu_0)}$ is still 
positive so that $\Gamma(M_{N(\mu_0)})$ is not empty, and the terms 
containing the phase are bounded by
\begin{equation}
\mu_0^{N(\mu_0)}M_{N(\mu_0)} = \frac{M_0}{2^{N(\mu_0)}} =
M_0 \e^{-\ln 2 \brak{D_0/d\mu_0}} = M_0 \e^{-1/C\mu_0}.
\label{pth22}
\end{equation}

Finally, we also see that the terms defining the successive changes of 
variables, i.e. $\mu_0^{N+1}f_N$, where $f_N\sim M_N$, add up to a 
geometrical series converging towards a term of order $\mu_0$.

 
\subsection{Conservative case}

 
It is easy to check that if (\ref{th1}) preserves the measure 
$c(I)\rho(\ph,I,\mu)d\ph dI$, $\rho(\ph,I,0)=1$, then (\ref{pth1}) 
preserves the measure $c(I)\rho_N(\ph,I,\mu)d\ph dI$, $\rho_N(\ph,I,0)=1$. 
As long as (\ref{pth1}) is $C^1$ and $c\in C^2$, we can apply the 

{\bf Lemma 2}
{\it
If (\ref{pth1}) preserves the measure $c(I)\rho d\ph dI$, then 
$\Theta_N\im0=\Order{\mu^{N+1}}$ and $\pd{\ph}\rho=\Order{\mu^{N+1}}$.
}

{\bf Proof:}
We assume by induction that $\Theta_N c(I)= \mu^p \ThNbar$ 
and $\rho(\ph,I,\mu) = \rho_0(I,\mu) + \mu^{p+1}\rho_1(\ph,I,\mu)$, 
which is clearly true for $p=0$. We have
\begin{eqnarray}
\frac{c(I_1)\rho\Pim1}{c(I_0)\rho\pim0} & = & 
\det\matrix22{1+\mu^{N+2}c(I_0)\pd{\ph}\alpha_N}{\Order{\mu}}
{\Order{\mu^{N+2}}}{1+\mu^{p+2}\pd{I}(c\ThNbar)+\mu^{N+2}\pd{I}(c^2\beta_N)}
\nonumber \\
 & = & 1 + \mu^{p+2}\pd{I}\brak{c(I_0)\ThNbar(I_0,0)}
 + \mu^{N+2}\parth{c\pd{\ph}\alpha_N+\pd{I}\brak{c^2\beta_N}} +
\nonumber \\
 &  & + \Order{\mu^{p+3}}+\Order{\mu^{N+3}}.
\label{pth23}
\end{eqnarray}

On the other hand, making expansions, we obtain
\begin{eqnarray}
\frac{c(I_1)\rho\Pim1}{c(I_0)\rho\pim0} & = & 
1 + \mu^{p+2}\brak{c(I_0)\W0\pd{\ph}\rho_1(\ph_0,I_0,0)+c'(I_0)\ThNbar(I_0,0)}
+ \nonumber \\
 &  & + \mu^{N+2}c(I_0)c'(I_0)\beta_N(\ph_0,I_0,0) 
+ \Order{\mu^{p+3}}+\Order{\mu^{N+3}}.
 \label{pth24}
\end{eqnarray}

Comparing (\ref{pth23}) and (\ref{pth24}), we obtain for $p<N$
\begin{equation}
\W0\pd{\ph}\rho_1(\ph_0,I_0,0) = \pd{I}\ThNbar(I_0,0).
\label{pth25}
\end{equation}

The left-hand side is a periodic function of $\ph$ with average $0$. 
Hence, the right-hand side must vanish. This implies $\pd{\ph}\rho_1=0$ 
and $\ThNbar=\mbox{const}$. But this constant must be zero since $\ThNbar$ 
is divisible by $c(I)$. This proves $p=N$ by induction. For $p=N$, one 
obtains again that $\pd{I}\ThNbar$ is a  periodic function of $\ph$ with 
average $0$, which proves the lemma. \qed

In the $C^k$ case, we first do $k-1$ changes of variables, so that the map 
takes the form (\ref{pth1}) with $N=k-1$ and is $C^1$. Then we can apply 
the above lemma, and finally we do the change of variables (\ref{pth2})
with $N=k$, which proves (\ref{th5}).

In the analytic case, after $N(\mu_0)$ changes of variables, the lemma 
implies that we can write $\Theta_N(I,\mu)=\mu^N\hat{\Theta}_N(I,\mu)$, 
and Cauchy's inequality gives
\begin{equation}
\abs{\hat{\Theta}_N(I,\mu)}\leq\frac{\overline{W}}{(\mu_0-\abs{\mu})^N}.
\label{pth26}
\end{equation}

For $\mu_1=\mu_0/3$, we have
\begin{equation}
\abs{\mu_1^N \hat{\Theta}_N} \leq \frac{\overline{W}}{2^N},\;\;\;
\abs{\mu_1^N \beta_N} \leq \frac{M_0}{6^N},
\label{pth27}
\end{equation}
which proves (\ref{th3}).\qed

 
\section{Conclusion}
\label{sec_concl}


In this work on classical billiards in plane domains, we found that some 
properties of bouncing maps, which are known in the Euclidean case, can 
be generalized to the situation where a magnetic field is applied 
perpendicularly to the plane. Exact expressions for the Jacobian matrix 
and a generating function help to find periodic orbits, analyze their
stability and compute bifurcation values. Perturbative calculations 
are improved.

Some aspects of the behaviour of billiards in  convex domains are well 
understood: if the boundary is sufficiently smooth, existence of whispering 
gallery modes prevents ergodicity, and when the magnetic field $B$ goes 
to infinity, the bouncing map behaves like a perturbed integrable map.
We were able to construct quasi-invariants, which are conserved for a time
of order $B^{k-3}$ if the boundary is $C^k$, and of order $\e^B$ when the 
boundary is analytic. Some of these properties remain valid for a more 
general class of billiards, the boundary of which is not convex, but 
has a bounded curvature.

Chaotic behaviour seems to be created by two different mechanisms:
non-linearity (responsible for instance of chaotic components near 
separatrices) and singularities (due to non-smooth boundaries or 
discontinuities by tangency). Non-linearity alone is not sufficient 
to make a magnetic billiard ergodic, since we showed that there always 
exist invariant curves if the boundary is $C^6$ and tangencies are 
impossible. This implies in particular that billiards with boundaries 
of negative curvature are not necessarily very chaotic.
In any case where the billiard is of the mixed type, the really 
challenging open problem remains to prove the existence of a chaotic 
component of positive measure.

Finally, we discussed two examples of billiards fulfilling necessary 
conditions for ergodicity (i.e. their bouncing maps have singularities).
Our study of symmetric periodic orbits in the square lead us to the conjecture 
that ergodic and mixed dynamics alternate when the magnetic field increases.
However, only the first value of supposed ergodicity, 
$\mu = \frac{\sqrt{5}}{2}$, lies in the region $\mu>\half$, where no trivially 
integrable component of phase space exists.
The scattering billiard outside two circles shows strongly chaotic 
dynamics, but still possesses elliptic orbits for some values of the 
parameters.

 
\section*{Acknowledgments}
\label{sec_ack}


We benefitted from talks with prof. D. Sz\'asz, C. Rouvinez, T. Dagaeff, 
K. Rezakhanlou and several other billiard players. This work is supported
by the Fonds National Suisse de la Recherche Scientifique.


\appendix


\section{Proof of proposition 1}
\label{app1}


We consider the arc of trajectory in figure \ref{fig_arc}. If
\begin{equation}
 \alpha = \arg \brak{\parth{X(s_1)-X(s_0)} + \icx \parth{Y(s_1) - Y(s_0)}}
\label{A1}
\end{equation}
is the angle between the $x$-axis and $P_0P_1$, then we have
\begin{eqnarray}
 \theta_0 &=& \alpha - \tau_0 - \chi, \nonumber \\
 \theta_1 &=& \tau_1 - \alpha - \chi.
\label{A2}
\end{eqnarray}
Now, using (\ref{map1b}) and $\ell^2 = (X(s_1)-X(s_0))^2 + (Y(s_1) - Y(s_0))^2$,
we find
\begin{eqnarray}
 \dpart{\alpha}{s_0} &=& \frac{1}{\ell^2}
 \brak{\parth{Y(s_1) - Y(s_0)}X'(s_0) - \parth{X(s_1) - X(s_0)}Y'(s_0)}\nonumber \\
 &=& \frac{\sin\alpha\cos\tau_0 - \cos\alpha\sin\tau_0}{\ell} 
 = \frac{\sin(\theta_0+\chi)}{\ell}, \nonumber \\
 \dpart{\tau_0}{s_0} &=& \kappa(s_0) = \kappa_0, \nonumber \\
 \dpart{\chi}{s_0} &=& \frac{1}{\cos\chi} \frac{1}{2\mu} \dpart{\ell}{s_0}
 = -\frac{\cos(\theta_0+\chi)}{2\mu\cos\chi}
 = -\frac{\cos(\theta_0+\chi)\sin\chi}{\ell\cos\chi}, 
\label{A3}
\end{eqnarray}
where the last equality comes from the zero-field generating property of $\ell$:
\begin{eqnarray}
 \dpart{\ell}{s_0} &=& \frac{1}{\ell}
 \brak{-X'(s_0)\parth{X(s_1) - X(s_0)} - Y'(s_0)\parth{Y(s_1) - Y(s_0)}}\nonumber \\
 &=& -\cos\alpha\cos\tau_0 - \sin\alpha\sin\tau_0 = -\cos(\theta_0+\chi).
\label{A4}
\end{eqnarray}
Collecting terms, we get
\begin{equation}
 \dpart{\theta_0}{s_0} = \frac{\sin(\theta_0+2\chi)}{\ell\cos\chi} - \kappa_0, \phantom{MM}
 \dpart{\theta_1}{s_0} = -\frac{\sin\theta_0}{\ell\cos\chi}.
\label{A5}
\end{equation}
Similarly, we find
\begin{equation}
 \dpart{\theta_0}{s_1} = \frac{\sin\theta_1}{\ell\cos\chi}, \phantom{MM}
 \dpart{\theta_1}{s_1} = -\frac{\sin(\theta_1+2\chi)}{\ell\cos\chi} + \kappa_1.
\label{A6}
\end{equation}

These quantities give $d\theta_0$ and $d\theta_1$ in function of $ds_0$ 
and $ds_1$. Solving a linear system, we can express $ds_1$ and $d\theta_1$
as functions of $ds_0$ and $d\theta_0$. Finally, using 
$du = \sin\theta d\theta$, we obtain the equations (\ref{map2}).\qed


\section{Proof of proposition 2}
\label{app2}


Taking the derivative of (\ref{gen5}) and using (\ref{map1b}), we get
\begin{equation}
 \dpart{B_\mu}{\ell} = \sqrt{1-\frac{\ell^2}{4\mu^2}} = \cos\chi.
\label{B1}
\end{equation}
From simple geometry, we obtain
\begin{equation}
 \dpart{A}{s_0} = -\half \ell \sin(\theta_0+\chi),
\label{B2}
\end{equation}
so that the derivative of (\ref{gen4}) is
\begin{equation}
 \dpart{G}{s_0}
 = \frac{1}{\mu} \dpart{A}{s_0} + \dpart{B_\mu}{\ell}\dpart{\ell}{s_0} 
 = -\sin\chi \sin(\theta_0+\chi) - \cos\chi \cos(\theta_0+\chi)
 = u_0,
\label{B3}
\end{equation}
where we have used (\ref{A4}) and (\ref{map1b}) again. We proceed in a
similar way for $s_1$.\qed

\begin{figure}[t]
\centerline{\psfig{figure=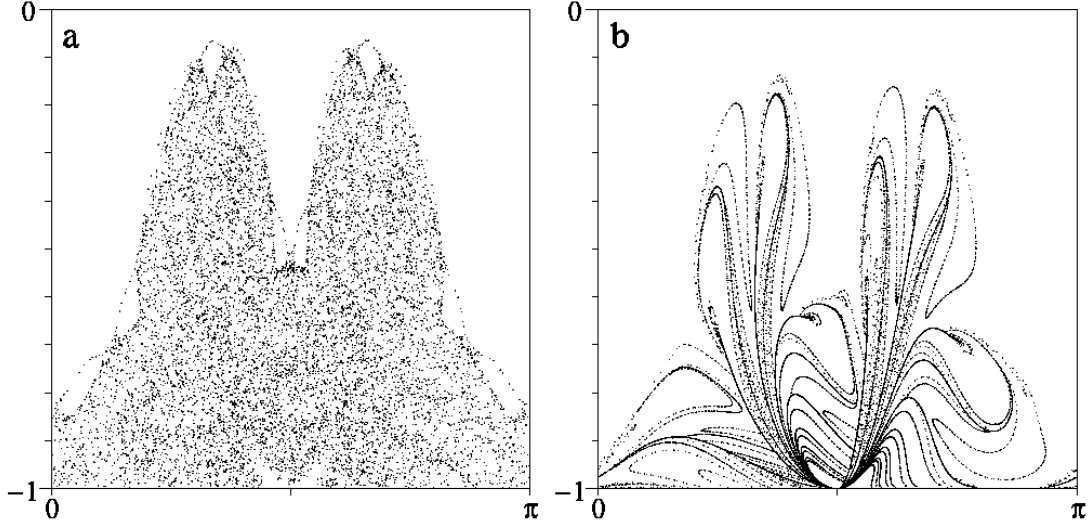,height=70mm,clip=t}}
\caption{Behaviour near $u=-1$ of the billiard in an ellipse when 
 	 $\rmin<\mu<\rmax$
	 ($\mu = 1$, $\lambda = 2$):
	 (a) chaotic orbit near $u=-1$, (b) the preimages $\D_n$, 
	 $n=1\ldots 15$ of the line $u=-1$.}
\label{fig_disc}
\end{figure}


\section{Discontinuities of the map due to tangencies}
\label{app4}


The bouncing map is discontinuous either in the corners of the boundary
($s \in E_1$), or if the trajectory becomes tangent to the boundary.
As can be seen in fig.\ref{fig_ellipse}b, such a tangency means that 
there are two arbitrarily close initial conditions $z_1, z_2$ such 
that $Tz_2$ is far away from $Tz_1$ but close to $T^2z_1$. Then there 
exists a point $z^*$ between $z_1$ and $z_2$ such that $Tz^* = (s^*, -1)$ 
(or $(s^*, +1)$ if $\dQ$ is not convex).

Let $\C = [0, \Lb) \times \set{-1, +1}$. It can be written as the union of 
3 disjoint sets:

\begin{itemize}
 \item	the set $\C_0$ of $z \in \C$ such that $T^{-1}z$ does not exist;

 \item	the set $\C_1$ of $z \in \C$ such that $T^{-1}z \in \C$;

 \item	the set $\C_2$ of $z \in \C$ such that $T^{-1}z \not\in \C$.
\end{itemize} 

The set $\D$ of lines of discontinuity of $T$ is $T^{-1}\C_2$.

To construct the sets $\C_i$, we first look for circles of radius $\mu$ 
which are {\em inscribed} in $Q$, i.e. circles contained in $\overline{Q}$ 
and tangent to $\dQ$ at two points or more. Let $\Ebar$ be the 
set of abscissas $\sbar_i$ of these contact points. Let $\Etilde$ be 
set of abscissas $\stilde_j$ such that $\abs{\rho(\stilde_j)} = \mu$.
Then $\C_0$, $\C_1$ and $\C_2$ are delimited by points with abscissa in 
$\Ebar$, $\Etilde$ and $E_2$ (see \cite{BeRo} for a more geometric 
interpretation of these sets).

We illustrate this construction in the case of an elliptic boundary
(\ref{gen6}), using $\varphi$ instead of $s$. Here we take 
$\C = [0, \Lb) \times \set{-1}$. Inscribed circles exist if 
$\rmin = \lambda^{-1} < \mu \leq 1$, and their contact points are 
solutions of
\begin{equation}
 \sin^2 \phb_i = \frac{(\lambda\mu)^2-1}{\lambda^2-1}.
\end{equation}
If $\rmin = \lambda^{-1} < \mu \leq \rmax = \lambda^2$, there are 
points $\pht_j$, given by
\begin{equation}
 \sin^2 \pht_j = \frac{(\lambda\mu)^{2/3}-1}{\lambda^2-1}.
\end{equation}
If they exist, the solutions can be ordered 
$0\leq\pht_1\leq\phb_1\leq\phb_2\leq\pht_2\leq
\pht_3\leq\phb_3\leq\phb_4\leq\pht_4\leq 2\pi$. They satisfy
\begin{eqnarray}
 T^{-1}(\phb_i,-1) = (\phb_{5-i}, -1) &\mbox{for}& i=1,2,3,4, \\
 \limit{\varphi}{\pht_j+} T^{-1}(\varphi,-1) = (\pht_j, -1) &\mbox{for}& j=1,3.
\end{eqnarray}

Table \ref{tab1} shows the resulting subsets of $\C$. Following ideas in 
\cite{DaRo}, one can look at the successive preimages $\D_n=T^{-n}\C_2$ 
of the lines of discontinuity. Numerical simulations show that they 
densely fill the stochastic layer near $u=-1$ (see fig.\ref{fig_disc}).

\begin{table}[h]
 \centering
 \small
 \caption[]{The sets $\C_0$, $\C_1$ and $\C_2$ for an elliptic boundary.}
 \label{tab1}
 \begin{tabular}{cccc}
  \noalign{\vskip 3pt}
  \hline
  \noalign{\vskip 3pt}
  $\mu \in$ 	& $\C_0$ 	& $\C_1$ 	& $\C_2$ 	\\
  \noalign{\vskip 3pt}
  \hline
  \noalign{\vskip 3pt}
  $(0,\rmin]$	& $\emptyset$	& $\C$		& $\emptyset$	\\
  \noalign{\vskip 3pt}
  $(\rmin,1]$	& $(\pht_2,\pht_3] \cup (\pht_4,\pht_1]$ 
		& $[\phb_1,\phb_2] \cup [\phb_3,\phb_4]$
		& $(\pht_1,\phb_1) \cup (\phb_2,\pht_2] \cup$	\\
  		& $\times \set{-1}$ 
		& $\times \set{-1}$
		& $(\pht_3,\phb_3) \cup (\phb_4,\pht_4] \times \set{-1}$\\
  \noalign{\vskip 3pt}
  $(1,\rmax)$	& $(\pht_2,\pht_3] \cup (\pht_4,\pht_1]$ 
		& $\emptyset$	
		& $(\pht_1,\pht_2] \cup (\pht_3,\pht_4]$	\\
  		& $\times \set{-1}$ 
		& 	& $\times \set{-1}$			\\
  \noalign{\vskip 3pt}
  $[\rmax,\infty)$	& $\C$	& $\emptyset$	& $\emptyset$	\\ 
  \noalign{\vskip 3pt}
  \hline
  \noalign{\vskip 3pt}
 \end{tabular}
\end{table}


\section{Proof of proposition 3}
\label{app5}


The following geometrical lemma describes some properties of convex 
plane curves (which seem quite obvious if one draws a picture).

\begin{figure}[t]
 \centerline{\psfig{figure=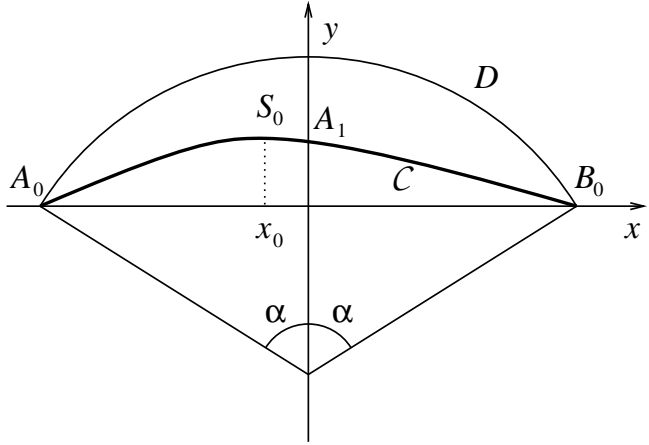,height=55mm}}
 \caption{Geometry of the circular segment of lemma 3.}
\label{fig_segcirc}
\end{figure}

\propsep
{\bf Lemma 3}
{\it 
Let $D$ be a circular segment of angle $2\alpha\in(0,\pi]$ and radius $1$.
Let $\C$ be a strictly convex $C^2$ curve with extremities on the vertices
$A_0, B_0$ of $D$, and making acute or right angles with the chord $A_0B_0$
(see fig.\ref{fig_segcirc}). Then, if the curvature of $\C$ is everywhere less 
than $1$, it is entirely contained in $D$ and shorter than $2\alpha$.}

{\bf Proof:}
We introduce coordinate axes as in figure \ref{fig_segcirc}. For 
$-\sin\alpha<x<\sin\alpha$,
$\C$ can be described by a function $y(x)$, and the curvature is
\begin{equation}
\kappa(x) = -\frac{y''(x)}{(1+y'(x))^{3/2}} \in (0,1).
\label{L1}
\end{equation}
This equation can be integrated between the abscissa $x_0$ of the maximum 
$S_0$ and $x$:
\begin{eqnarray}
y'(x) &=& -\frac{k(x)}{\sqrt{1-k(x)^2}} \nonumber \\
k(x) &=& \int_{x_0}^x \kappa(x_1) dx_1, \; \abs{k(x)} < \abs{x-x_0}.
\label{L2}
\end{eqnarray}
For $x_0-1<x<x_0+1$, this implies
\begin{equation}
\brak{y(x)-(y(x_0)-1)}^2 + (x-x_0)^2 \geq 1,
\label{L3}
\end{equation}
and hence $\C$ lies above the circle of radius 1 tangent to it at its summit
$S_0$. Now assume by contradiction that $\C$ lies outside $D$ between two 
points $P$ and $P'$. Let $S$ be the point of $\C$ furthest away from $PP'$. 
The circle of unit radius tangent to $\C$ at $S$ cannot intersect $D$
twice between $P$ and $P'$, which contradicts (\ref{L3}).

If $P$ and $P'$ are two points of $\C$, we denote by $\lenarc{PP'}$ the length 
of $\C$ between these points. If $P(x)=(x,y(x))$, then for $x_0-1<x<x_0+1$, 
\begin{equation}
\lenarc{S_0P(x)}=\int_{x_0}^x \sqrt{1+y'(x_1)^2} dx_1 \leq \Asin \abs{x-x_0}.
\label{L4}
\end{equation}
 Assume that $x_0\leq 0$. Then 
$\lenarc{A_0A_1}\leq\Asin(x_0+\sin\alpha)+\Asin(-x_0)\leq\alpha$ by 
convexity of $\Asin$. Considering the circular segment of vertices 
$A_1$ and $B_0$, one obtains that $\lenarc{A_1B_0}$ can be divided into 
two parts, one of which is bounded by $\frac{\alpha}{2}$. Repeating this 
procedure, we see that $\C$ can be divided into pieces whose lengths are 
bounded by a geometrical series, converging towards $2\alpha$. \qed

\propsep
{\bf Corollary}
{\it
Let $\C$ be a $C^2$ plane convex curve, whose curvature satisfies 
$0<\frac{1}{\rmax}\leq\kappa(s)\leq\frac{1}{\rmin}<\infty$.
Let $\Gamma$ be a circle of radius $\mu<\rmin$. Then
\begin{enumerate}

\item $\Gamma$ can be tangent to $\C$ at one point at most.

\item $\Gamma$ can intersect $\C$ at two points at most.

Assume that $\Gamma$ intersects $\C$ at $P$ and $P'$ and call 
$\C_0 = \C \cap {\rm Int}\Gamma$.

\item $\C_0$ is shorter than $\pi\mu$.

\item The angles between $PP'$ and $\C_0$ are acute.

\end{enumerate}
}
{\bf Proof:}
Assuming that 1. or 2. are false, one can construct a counterexample 
of the lemma (one may have to translate $\Gamma$). 3. and 4. are 
obvious if one considers all the circles of radius $\mu$ 
containing $P$ and $P'$.\qed

\propsep

\begin{figure}
 \centerline{\psfig{figure=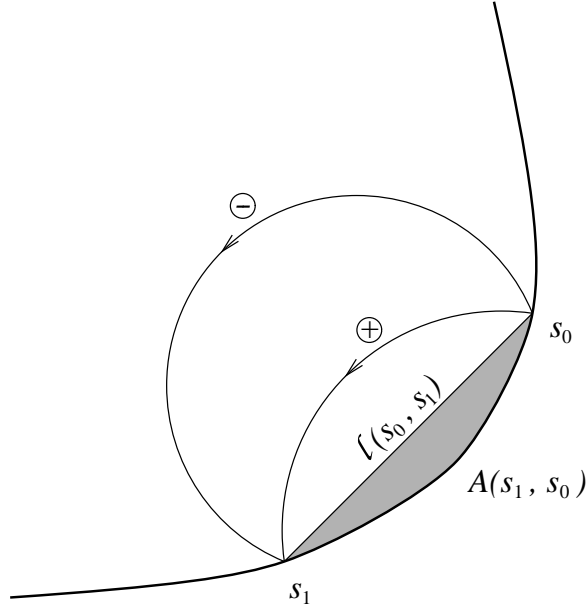,height=80mm}}
 \caption{Geometry of the trajectories for a convex billiard in strong 
 magnetic field.}
\label{fig_strfield}
\end{figure}

We now proceed to the proof of the proposition.

Choose two points $P$ and $P'$ on $\dQ$. They can be connected by 
two, one or no arc of trajectory, depending on whether their distance 
is smaller than, equal to, or larger than $2\mu$. In each of the 
first two cases, we call $s_0$ and $s_1$ their abscissas, as in figure
\ref{fig_strfield}.
We define
\begin{equation}
\sigma = \frac{(s_0-s_1)\pmod{\Lb}}{\mu}.
\label{pp3_1}
\end{equation}
The corollary and our sign conventions imply that $0<\sigma<\pi$.
Moreover, the arcs cannot intersect $\dQ$ at a point different 
from $P$ or $P'$.

Following the proof of proposition 2, it is easy to show that the 
two trajectories can be described by the generating functions
\begin{equation}
G^{\pm}(s_0,s_1) = -\frac{1}{\mu} A(s_1,s_0) 
\pm \mu b\parth{\frac{\ell(s_0,s_1)}{2\mu}},
\label{pp3_2}
\end{equation}
in the sense that $dG^{\pm} = u_0^{\pm}ds_0 - u_1^{\pm}ds_1$.
Here, $\ell\in(0,2\mu]$ and $A$ are defined in fig.\ref{fig_strfield}, and 
\begin{eqnarray}
b(x) & = & \Asin x + x \sqrt{1-x^2} \nonumber \\
b'(x) & = & 2 \sqrt{1-x^2}.
\label{pp3_3}
\end{eqnarray}

The functions $A$ and $\ell$ are directly related to the shape 
of $\dQ$, that we describe using the function $\tau(s)$, 
which has the properties
\begin{enumerate}

\item $\tau(s+\Lb) = \tau(s)+2\pi$,

\item $\tau'(s) = \kappa(s) \in [\frac{1}{\rmax},\frac{1}{\rmin}]$,

\item $\ds\int_0^{\Lb} \math{e}^{\icx \tau(s)} ds = 0$.

\end{enumerate}

If $\vec{\ell}(s_0,s_1)$ is the vector connecting the points of 
abscissas $s_0$ and $s_1$, we have
\begin{eqnarray}
\vec{\ell}(s_0,s_1) & = & \int_{s_0}^{s_1} \vec{t}(s)ds
\nonumber \\
\ell(s_0,s_1)^2 & = & \int_{s_0}^{s_1} ds \int_{s_0}^{s_1} ds'
\pscal{\vec{t}(s)}{\vec{t}(s')} = 
\int_{s_0}^{s_1} ds \int_{s_0}^{s_1} ds' \cos\brak{\tau(s)-\tau(s')}
\nonumber \\
A(s_0,s_1) & = & \int_{s_0}^{s_1} ds
\half \abs{\vec{\ell}(s_0,s) \wedge \vec{t}(s)} = \half
\int_{s_0}^{s_1} ds \int_{s_0}^{s} ds' \sin\brak{\tau(s)-\tau(s')}.
\label{pp3_4}
\end{eqnarray}
Carrying out the change of variables $s = s_0 + \mu\sigma t$,
we obtain $\ell(s_0,s_1)^2 = \mu^2\sigma^2 I(s_0, \mu\sigma)$, 
$A(s_1,s_0) = \half\mu^2\sigma^2 I'(s_0, \mu\sigma)$, where
\begin{eqnarray}
I(s_0, \mu\sigma) & = & \int_{-1}^0 dt \int_{-1}^0 dt' 
\cos T(s_0, \mu\sigma t, \mu\sigma t')
\nonumber \\
I'(s_0, \mu\sigma) & = & \int_{-1}^0 dt \int_{-1}^t dt' 
\sin T(s_0, \mu\sigma t, \mu\sigma t')
\nonumber \\
T(s_0, \mu\sigma t, \mu\sigma t') & = & 
\tau(s_0+\mu\sigma t) - \tau(s_0+\mu\sigma t').
\label{pp3_5}
\end{eqnarray}
If the boundary is $C^2$, one gets $T = \Order{\mu\sigma}$, 
$I = 1+\Order{\mu\sigma}$, $I' = \Order{\mu\sigma}$, 
and thus $G^{\pm}(s_0,s_1,\mu) = \mu g^{\pm}(\sigma, s_0,\mu)$,
where $g^{\pm}(\sigma, s_0,0) = \pm b\parth{\frac{\sigma}{2}}$
does not depend on $s_0$.
From this fact, we can already guess the structure of the map.
Indeed, $u_0^\pm = \pd{\sigma}g^\pm + \mu\pd{s_0}g^\pm,
u_1^\pm = \pd{\sigma}g^\pm$, and thus 
$u_1^\pm = u_0^\pm +  \mu\pd{s_0}g^\pm = u_0^\pm + \Order{\mu^2}$.

However, $G$ is not sufficiently smooth around $\ell=2\mu$ to be 
expanded. We proceed in a slightly different way: taking derivatives 
of (\ref{pp3_4}), we obtain
\begin{equation}
\dpart{\ell}{s_0}=\frac{\mu}{\ell}\sigma J,\;\;\;
\dpart{A}{s_0}=\frac{\mu}{2}\sigma K,\;\;\;
\dpart{A}{s_1}=-\frac{\mu}{2}\sigma K',
\label{pp3_6}
\end{equation}
where
\begin{eqnarray}
J(s_0, \mu\sigma) & = & \int_{-1}^0 dt \cos T(s_0, 0, \mu\sigma t)
\nonumber \\
K(s_0, \mu\sigma) & = & \int_{-1}^0 dt \sin T(s_0, 0, \mu\sigma t)
\nonumber \\
K'(s_0, \mu\sigma) & = & \int_{-1}^0 dt \cos T(s_0, \mu\sigma t, -\mu\sigma).
\label{pp3_7}
\end{eqnarray}
Note that if the boundary is $C^k$, the above integrals are all $C^{k-1}$.
Differentiating (\ref{pp3_2}), we get
\[
u_0^\pm = \dpart{G^\pm}{s_0} = 
\pm\sqrt{1-\frac{\ell^2}{4\mu^2}}\;
\dpart{\ell}{s_0}-\frac{1}{\mu}\dpart{A}{s_0}
=\pm\sqrt{1-\frac{1}{4}\sigma^2 I}\;
 \frac{\mu}{\ell}\sigma J-\half \sigma K
 \]
\begin{eqnarray}
\parth{u_0+\half \sigma K}^2 & = & 
\parth{1-\frac{1}{4}\sigma^2 I}\frac{J^2}{I}
\nonumber \\
4I(1-u_0^2) = 4I\sin^2\th_0 & = & 
\sigma^2(K^2+J^2)I-4\sigma\cos\th_0KI+4(I-J^2)
\nonumber \\
 & = & \Phi_0(\sigma,\mu,s_0,\th_0).
\label{pp3_8}
\end{eqnarray}

When the boundary is $C^3$, we have $I=1+\Order{\mu^2\sigma^2}$,
$J=1+\Order{\mu^2\sigma^2}$ and 
$K=\half\kappa(s_0)\mu\sigma+\Order{\mu^2\sigma^2}$, so that
$\Phi_0=\sigma^2\brak{1-2\cos\th_0\kappa(s_0)\mu+\Order{\mu^2}}$.

If we write $\sigma=2\sin\th_0\eta$, then
\begin{equation}
0=\Phi(\eta,\mu,s_o,\th_0)=
\evaluate{\brak{\sqrt{I} - \eta \sqrt{\frac{\Phi_0}{\sigma^2}}\,}}
{\sigma=2\sin\th_0\eta}
= 1-\eta\brak{1-\cos\th_0\kappa(s_0)\mu+\Order{\mu^2}}.
\label{pp3_9}
\end{equation}

The function $\Phi$ has the properties 
$\Phi(\eta_0,0,s_0,\th_0) = 0$, where $\eta_0=1$, and
$\pd{\eta}\Phi(\eta_0,0,s_0,\th_0) = -1$. Thus, the implicit
function theorem implies that in a neighborhood  of $\mu=0$, $\eta$
can be expressed as a $C^{k-1}$ function of  $\mu, s_0$ and $\th_0$.
This function can be constructed using  Newton's method: if
$\eta_0\equiv 1$ and $\eta_{n+1}(s_0,\th_0,\mu)  =
\eta_n(s_0,\th_0,\mu) + \Phi(\eta_n(s_0,\th_0,\mu),\mu,s_0,\th_0)$,
then  $\Phi(\eta_n,\mu,s_0,\th_0)=\Order{\mu^{n+1}}$. For $k=3$, we get
\begin{equation}
\eta(s_0,\th_0,\mu) = 1 + \cos\th_0\kappa(s_0)\mu + \Order{\mu^2}.
\label{pp3_10}
\end{equation}

The first equation of (\ref{prop3_1}) is obtained by using 
$s_1 = s_0-\mu\sigma = s_0-2\mu\sin\th_0\eta$, i.e.
\begin{equation}
a(s_0,\th_0,\mu) = \frac{2}{\mu}(1-\eta(s_0,\th_0,\mu))\in C^{k-2}.
\label{pp3_11}
\end{equation}
Note that if we had used the variable $u$ instead of $\th$, 
we would not have been able to apply the implicit function 
theorem when $u\rightarrow \pm 1$. Indeed, the map expressed in 
the variables $(u,s)$ contains the factor $\sqrt{1-u_0^2}$.

To obtain the second equation of (\ref{prop3_1}), we observe that 
(see (\ref{B2})):
\begin{eqnarray}
\th_1-\th_0 & = & (\th_1+\chi) - (\th_0+\chi) \nonumber \\
& = & \brak{\pi-\Asin\parth{\frac{2}{\ell}\dpart{A}{s_1}}}
- \brak{\pi-\Asin\parth{-\frac{2}{\ell}\dpart{A}{s_0}}}
\nonumber \\
& = & \Asin\parth{\frac{K}{\sqrt{I}}}
- \Asin\parth{\frac{K'}{\sqrt{I}}} = \Order{\mu^2\sigma^2},
\label{pp3_12}
\end{eqnarray}
where we have used point 4 of the corollary to choose the determination
of $\Asin$. By consequence,
\begin{equation}
b(s_0,\th_0,\mu) =
\evaluate{\frac{4\eta^2}{\mu^2\sigma^2}\brak{
\Asin\parth{\frac{K}{\sqrt{I}}}
- \Asin\parth{\frac{K'}{\sqrt{I}}}}}{\sigma=2\sin\theta_0\eta}
\in C^{k-3}. 
\label{pp3_13} 
\end{equation}

Higher orders of the expansions of $a$ and $b$ can be obtained by
computing expansions of $I,J,K,K'$ and $\eta$ up to the desired order, 
and replacing them in (\ref{pp3_11}) and (\ref{pp3_13}).\qed

\pagebreak


\section{Symmetric orbits of period $4k$ in the square}
\label{app3}

\begin{figure}
 \centerline{\psfig{figure=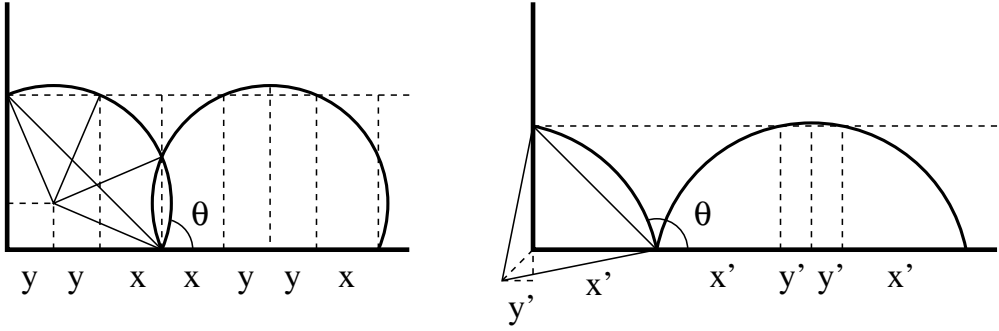,height=40mm}}
 \caption{Geometry of symmetric $4k$-periodic trajectories in the square.}
\label{fig_per4k}
\end{figure}

We want to study existence and stability of $4k$-periodic trajectories, similar 
to the one shown in figure \ref{fig_square}f, where $k = 2$.
As shown in figure \ref{fig_per4k}, such trajectories can be characterized
by numbers $x$, $y$ or $x'$, $y'$, satisfying
\begin{equation}
 \begin{array}{ll}
 \accol2{2kx+2(k+1)y=1}{2y^2+2xy+x^2=\mu^2,} \phantom{MM}&
 \accol2{0\leq y<\half}{0\leq 2y+x\leq\half,} 	 \\
						 \\
 \accol2{2kx'+2(k-1)y'=1}{2y'^2+2x'y'+x'^2=\mu^2,} &
 \accol2{0\leq y'}{0<x'\leq\half.}
 \end{array}
\label{C1}
 \end{equation}
These equations have 3 different kinds of solutions:
\begin{equation}
 \begin{array}{lcll}
  y_+&=&\ds\frac{1+R}{2(k^2+1)}, \phantom{MMM} &\mu_k^- \leq\mu< \ds\half, \\ 
  &&& \\
  y_-&=&\ds\frac{1-R}{2(k^2+1)},  &\mu_k^- \leq\mu\leq \ds\frac{1}{2k}, \\ 
  &&& \\
   y'&=&\ds\frac{-1+R}{2(k^2+1)}, &\ds\frac{1}{2k} \leq\mu< \ds\frac{1}{\sqrt{2}(k-1)}, \\ 
 \end{array}
\label{C2}
\end{equation}
where
\begin{equation}
 R = \ds\sqrt{1-(k^2+1)(1-4k^2\mu^2)}, \;\;\; \mu_k^- = \ds\frac{1}{2\sqrt{k^2+1}}.
\label{C3}
\end{equation}
Using (\ref{map2}), we find that
the stability matrix for $k$ bounces is
\begin{eqnarray}
 S_k&=&\matrix22{C}{\mu(1-C^2)}{-\frac{1}{\mu}}{C}
 \matrix22{1}{-2\mu C}{0}{1}^{k-1} 
 = \matrix22{C}{\mu (1-(2k-1)C^2)}{-\frac{1}{\mu}}{(2k-1)C}, \nonumber \\
 t&=&\half\Tr S_k = C k,
\label{C4}
\end{eqnarray}
where
\begin{equation}
 C = \cotg \theta = \frac{y}{y+x} = -\frac{y'}{y'+x'}.
\label{C5}
\end{equation}
Now, since
\begin{equation}
 S_{4k} = S_k^4, \;\;\; \half\Tr S_{4k} = 8t^4 - 8t^2 + 1,
\label{C6}
\end{equation}
the total orbit is hyperbolic if $\abs{t}>1$, parabolic if $\abs{t} = 0, 
\frac{1}{\sqrt{2}}, 1$, and elliptic otherwise. Applying this to (\ref{C4}), 
we find that $y_+$-orbits are hyperbolic as soon as $\mu>\mu_k^-$, 
$y_-$-orbits are never hyperbolic, and $y'$-orbits are hyperbolic if
$\mu>\mu_k^+$, where
\begin{equation}
 \mu_k^+ = \frac{\sqrt{k^2+1}}{2(k^2-1)}.
\label{C7}
\end{equation}
We have thus obtained that symmetric $4k$-periodic orbits may be stable 
only if $\mu_k^-<\mu<\mu_k^+$. These bounds have the properties
\begin{equation}
\begin{array}{rcccl}
 \mu_k^- &<&\ds\frac{1}{2k}&<&\mu_k^+, 	\nonumber \\
 &&&&					\nonumber \\
 \mu_{k-1}^-&>&\mu_k^+&&\mbox{if }\;k>2.
\label{C8}
\end{array}
\end{equation}
We see that when $\mu<\mu_k^-$, no $4k$-periodic orbit exists. At $\mu=\mu_k^-$,
a pair of such orbits with opposite stability appears in a saddle-node 
bifurcation. The stable one loses stability at $\mu=\mu_k^+$. 
Numerical simulations show that new stable orbits are created, but they 
quickly loose stability for some $\mu = \mu_k^*$. 
For $k>2$, this happens long before a $4(k-1)$-periodic orbit appears at 
$\mu=\mu_{k-1}^-$, and since no other stable orbits can be found in the 
interval, we are lead to the conjecture that the billiard in a square is 
ergodic when $\mu_k^* \leq \mu \leq \mu_{k-1}^-$.


\section{Elliptic orbits of period 6 outside 2 circles}
\label{app6}


We want to show that the billiard outside 2 circles described in section
\ref{sec_neg} possesses elliptic orbits for some values of the 
parameters.

The trajectory depicted in figure \ref{fig_circles}b for $\mu=\lambda=1$ 
turns out to be linearly marginally stable for these values of the 
parameters if we apply (\ref{map2}). Thus we have to analyse its 
stability for nearby value of $\mu$ and $\lambda$.

\begin{figure}
 \centerline{\psfig{figure=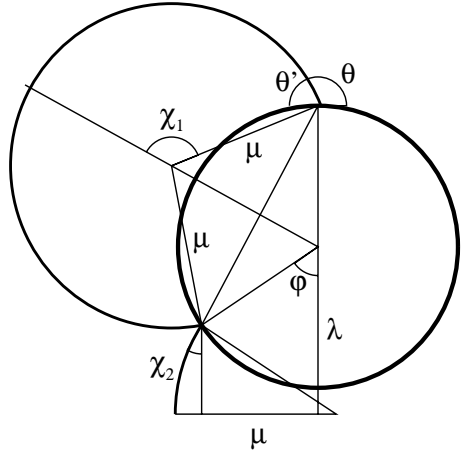,height=60mm}}
 \caption{Geometry of period 6 trajectories outside 2 circles. We show a 
	 quarter of the trajectory, which is symmetric with respect to the 
	 coordinate axes.}
\label{fig_period6}
\end{figure}

The trajectory can be characterized by two angles $\th'$ and $\ph$, as 
in figure \ref{fig_period6}. The other angles are then given by 
$\chi_1 = \frac{\pi}{2} + \th' - \frac{\ph}{2}$, 
$\chi_2 = \th' + \ph  - \frac{\pi}{2}$.
Using the relations $\mu\sin\chi_1 = \sin\parth{\frac{\pi-\ph}{2}}$ 
and $\cos\ph + \mu\sin\chi_2 = \lambda$, we obtain the system
\begin{equation}
\accol2
{1+\cos\ph = \mu^2 (1+\cos(\ph-2\th'))}
{\cos\ph - \mu\cos(\ph+\th') = \lambda.}
\label{app6_1}
\end{equation}
For small values of $\eps=\mu-1$ and $\delta=\lambda-1$, it has 
the solution
\begin{equation}
\left\{
\begin{array}{rcl}
\th' &=& \ds\frac{\pi}{3} - \ds\frac{1}{\sqrt{3}}\eps + 
\ds\frac{1}{\sqrt{3}}\delta + \Order{2}
\phantom{\ds\frac{1}{\ds\frac{1}{2}}}\\ 

\ph &=& \ds\frac{\pi}{3} - \frac{4}{\sqrt{3}}\eps + 
\frac{1}{\sqrt{3}}\delta + \Order{2},
\end{array}
\right. 
\label{app6_2}
\end{equation}
where $\Order{2}$ stands for terms of order $\eps^2$, $\delta^2$, $\eps \delta$.
The Jacobian matrices $M_1$ and $M_2$ for the two types of bounces can now 
be computed using (\ref{map2}), with $\th_i=\pi-\th'$ and 
$\ell_i \cos\chi_i = \mu\sin(2\chi_i)$. The stability matrix of the orbit 
is given by $S_6 = (S_3)^2$, $S_3 = M_1^2 M_2$. We obtain
\begin{eqnarray}
t & = & \half\Tr S_3 = -1 - 12 \eps + 6 \delta + \Order{2}, \nonumber \\
\Rightarrow \half\Tr S_6 & = & 2t^2-1 = 1 + 48 \eps - 24 \delta + \Order{2}.
\end{eqnarray}
Hence, for small $\delta$ and $\eps$, the orbit is elliptic for 
$\delta > 2 \eps + \Order{\eps^2}$, i.e. for 
$\lambda > 2 \mu - 1 + \Order{(\mu-1)^2}$, 
in a neighbourhood of $\lambda=\mu=1$.

\pagebreak




\end{document}